\documentclass[10pt,journal,compsoc]{IEEEtran}
\ifCLASSOPTIONcompsoc
  \usepackage[nocompress]{cite}
\else
  \usepackage{cite}
\fi
\ifCLASSINFOpdf
  \usepackage[pdftex]{graphicx}
\else
\fi
\usepackage{array}
\newcolumntype{M}[1]{>{\centering\arraybackslash}m{#1}}

\ifCLASSOPTIONcompsoc
 \usepackage[caption=false,font=footnotesize,labelfont=sf,textfont=sf]{subfig}
\else
 \usepackage[caption=false,font=footnotesize]{subfig}
\fi
\usepackage{url}

\usepackage{xspace}

\usepackage{listings}
\makeatletter
\def\lst@makecaption{%
  \def\@captype{table}%
  \@makecaption
}
\makeatother

\usepackage{capt-of}

\begin{document}

%
\title{A Qualitative Study of Application-level Caching}
%
%
%
%

\author{Jhonny~Mertz
        and~Ingrid~Nunes
\IEEEcompsocitemizethanks{
\IEEEcompsocthanksitem J. Mertz is with the Departamento de Inform{\'a}tica Aplicada, Instituto de Inform{\'a}tica, Universidade Federal do Rio Grande do Sul, Brazil.\protect\\
E-mail: jmamertz@inf.ufrgs.br
\IEEEcompsocthanksitem I. Nunes is with the Departamento de Inform{\'a}tica Aplicada, Instituto de Inform{\'a}tica, Universidade Federal do Rio Grande do Sul, Brazil and TU Dortmund, Germany.\protect\\
E-mail: ingridnunes@inf.ufrgs.br
}
\thanks{Manuscript received April 19, 2005; revised August 26, 2015.}}

%
%

\markboth{Journal of \LaTeX\ Class Files,~Vol.~14, No.~8, August~2015}%
{Shell \MakeLowercase{\textit{et al.}}: Bare Demo of IEEEtran.cls for Computer Society Journals}
%



\IEEEtitleabstractindextext{%
\begin{abstract}
Latency and cost of Internet-based services are encouraging the use of application-level caching to continue satisfying users' demands, and improve the scalability and availability of origin servers. Despite its popularity, this level of caching involves the manual implementation by developers and is typically addressed in an \emph{ad-hoc} way, given that it depends on specific details of the application. As a result, application-level caching is a time-consuming and error-prone task, becoming a common source of bugs. Furthermore, it forces application developers to reason about a crosscutting concern, which is unrelated to the application business logic. In this paper, we present the results of a qualitative study of how developers handle caching logic in their web applications, which involved the investigation of ten software projects with different characteristics. The study we designed is based on comparative and interactive principles of grounded theory, and the analysis of our data allowed us to extract and understand how developers address cache-related concerns to improve performance and scalability of their web applications. Based on our analysis, we derived guidelines and patterns, which guide developers while designing, implementing and maintaining application-level caching, thus supporting developers in this challenging task that is crucial for enterprise web applications.
\end{abstract}

}

\maketitle

\IEEEdisplaynontitleabstractindextext

%
\IEEEpeerreviewmaketitle

\IEEEraisesectionheading{\section{Introduction}\label{sec:introduction}}
\IEEEPARstart{D}{ynamically} generated web content represents a significant portion of web requests. This content changes according to a user request and due to this, the rate at which dynamic documents are delivered is often orders of magnitudes slower than static documents~\cite{Bouchenak2006,Ravi2009}. Therefore, dynamic content generation and delivery place a significant burden on servers, often leading to performance bottlenecks~\cite{Bouchenak2006}. To continue satisfying users' demands and also reducing the workload on content providers, over the past years many optimization techniques have been proposed~\cite{Ravi2009}. The ubiquitous solution to this problem is some form of \emph{caching}, which can reduce the user perceived latency and Internet traffic, and improve the scalability and availability of origin servers.

Recently, latency and cost of Internet-based services are driving the proliferation of \emph{application-level caching}, which is placed on the server-side and typically uses a key-value in-memory store system to cache frequently accessed or expensive to compute data that remain not cached in other caching levels, lowering the load on database or services that are difficult to scale up~\cite{Atikoglu2012,Ports2010,DellaToffola2015}. Therefore, application-level caching has become a popular technique for enabling highly scalable web applications.

Despite its popularity, the implementation of application-level caching is not trivial and demands high effort, because it involves manual implementation by developers. Its design and maintenance involve four key challenging issues: determining \emph{how} to cache the selected data, \emph{what} data should be cached, \emph{when} the selected data should be cached or evicted, and \emph{where} the cached data should be placed and maintained. Although application-level caching is commonly being adopted, it is typically implemented in an \emph{ad hoc} way, and there are no available practical guidelines for developers to appropriately design, implement and maintain caching in their applications. To find caching best practices, developers can make use of conventional wisdom, consult development blogs, or simply search online for tips. Nevertheless, this empirical knowledge is unsupported by concrete data or theoretical foundations that demonstrate its effectiveness in practice. Despite there are existing tool-supported approaches that can help developers to implement caching with minimal impact on the application code~\cite{Ports2010,Gupta2011,DellaToffola2015}, they do not consider all the aforementioned issues and do not take application specificities into account, letting most part of the reasoning, as well as the integration with the tool, to developers.

Given this context, we in this paper present a qualitative study performed to identify patterns and guidelines to support developers while designing, implementing and maintaining application-level caching in their applications. Previous qualitative studies have been conducted to gain a greater understanding of the behavior of developers during specific software development tasks~\cite{Robillard2004,Sillito2008,Nadi2015,Borstler2016,Namoun2016}. Moreover, there is work that focused on proposing software development guidelines to help software developers while developing applications~\cite{Jorgensen2005,Juristo2007,Carvajal2013}. Our work is similar in nature, but focuses on the investigation of a development challenge that has not been previously addressed.

Besides helping developers while handling cache, the novel findings of this study provide insights to propose solutions to raise the abstraction level of caching or automate caching tasks, providing a better experience with caching for developers. The results of this study are thus informative and provide practical guidance for developers. It is important to note, however, that qualitative research is aimed at gaining a deep understanding of a particular target of study, rather than providing a general description of a large sample of a population.

In short, this paper provides three main contributions: (i) the design and results of a qualitative study that provides an in-depth analysis of how developers employ application-level caching in web-based applications; (ii) useful guidelines for modeling an application-level caching component; and (iii) patterns for caching design, implementation and maintenance to be adopted under different circumstances.

The remainder of this paper is organized as follows. In Section~\ref{sec:background} we give background and related work on application-level caching. We detail the performed study in Section~\ref{sec:studydesign}, then present and discuss its results in Section~\ref{sec:analysis}. We describe the proposed guidelines and patterns derived from this study in Section~\ref{sec:guidelinesandpatterns}. Finally, we discuss threats to the validity of our study in Section~\ref{sec:threats} and, in Section~\ref{sec:discussion-conclusion}, we conclude.

 

\section{Background and Related Work}
\label{sec:background}

Focused on server-side, caching of dynamic web content can be implemented at several locations across a web-based system architecture~\cite{Podlipnig2003}. Depending on where caching is implemented, different types of content can be stored, such as the final HTML page~\cite{Candan2001}, intermediate HTML or XML fragments~\cite{Ramaswamy2005,Li2006a,Guerrero2011}, database queries~\cite{Soundararajan2005,Amza2005,Baeza-Yate2007,Ma2014}, or database tables~\cite{Larson2004}. These alternatives are conceived out of application boundaries and can cache dynamic data automatically, which provide transparent caching components for developers. However, they do not take application specificities into account, providing good results in general but are less effective where complex logic and personalized web content are processed within the application~\cite{Wang2014}. Therefore, application-level caching is an appealing technique to improve performance, reduce workload and make the overall user experience more pleasurable by reducing communication delays of web-based systems.

\begin{figure}
    \centering
    \includegraphics[width=\linewidth]{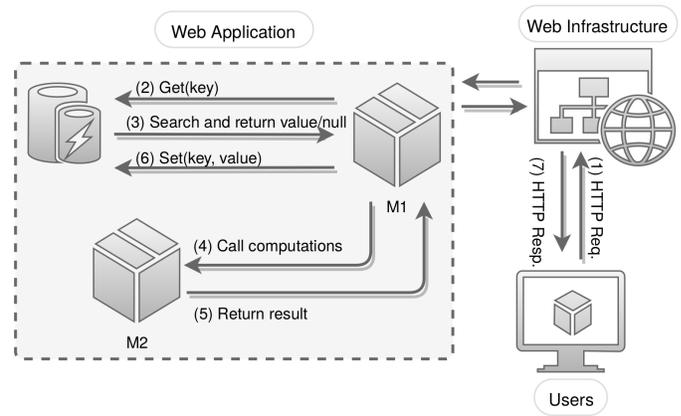}
    \caption{Using Application-level Caching to Decrease the Application Workload.}
    \label{fig:applicationcaching}
\end{figure}

Figure~\ref{fig:applicationcaching} illustrates an example where an application-level cache is used to lower the application workload. First, the application receives an HTTP request (Step 1) from the web infrastructure. This HTTP request is eventually treated by an application component called M1, which in turn depends on M2. However, M2 can be an expensive operation (i.e.\ request the database, call a service or process a large amount of data). Therefore, M1 implements a caching layer, which verifies whether the necessary processing is already in the cache before calling M2 (Step 2). Then, the cache component performs a look up for the requested data and returns either the cached result or a not found error (Step 3). If data is found, it means a cache \emph{hit} and M1 can continue its execution without calling M2. If, however, a not found error is returned, it means a cache \emph{miss} happened, then M2 needs to be invoked (Steps 4 and 5). The newly fetched result of M2 can be cached to serve future requests faster (Step 6) and eventually a response is sent to the user (Step 7). Furthermore, other caching layers can be deployed outside the application, at the web infrastructure.

The steps described in Figure~\ref{fig:applicationcaching} are those typically performed to implement cache at any level, and the key difference is the location where it is applied and its explicit manipulation by developers. Application-level cache allows caching at a granularity best suited to the application, providing a way to cache entire HTML pages, page fragments, database queries or even computed results; rather than laying in front of the application servers (e.g., a proxy-level cache) or between the application and the database (e.g., a query or database cache). For example, many websites have highly-personalized content, thus rendering whole-page web caches is mostly useless; application-level caches can be used to separate shared content from customized content, and then the shared content can be cached and reused among users~\cite{Ports2010}. Memcached~\cite{Memcached2016} and Redis~\cite{Redis2016} are popular solutions and are a critical web infrastructure component for some big players such as Amazon, Facebook, LinkedIn, Twitter, Wikipedia, and YouTube~\cite{Waddington2013}.

As already discussed in the introduction, the development of application-level caching involves four key issues: determining \emph{how}, \emph{what}, \emph{when}, and \emph{where} to cache data. The main problem is that solutions for all of these issues usually depend on application-specific details, and are manually designed and implemented by developers, as shown in the example presented in Listing~\ref{code:appCachingExample}. In this example, we assume that an e-commerce application where a class named \texttt{ProductsRepository} is responsible for loading products from database. To reduce database workload, caching logic is inserted in this class within the methods \texttt{getProducts}, \texttt{updateProduct}, and \texttt{deleteProduct}, which retrieves, updates and deletes content from the database, respectively.

\begin{lstlisting}[frame=single,caption=Example of Application-level Caching Implementation.,label=code:appCachingExample,float]
public class ProductsRepository {
  public List<Product> getProducts() {
    Cache cache = Cache.getInstance();
    List<Product> products = 
      (List<Product>) cache.get("products");
    if (products == null) {
        products = getProductsFromDB();
        cache.put("products", products);
    }
    return products;
  }
  public void updateProduct(Product product) {
    Cache.getInstance().delete("products");
    updateProductIntoDB(product);
  }
  public void deleteProduct(Product product) {
    Cache.getInstance().delete("products");
    deleteProductFromDB(product);
  }
}
\end{lstlisting}

The first issue is related to the \emph{cache-aside} implementation and the fact that the cache system and the underlying source of data are not aware of each other (as illustrated in Figure~\ref{fig:applicationcaching}). The cache system itself is a passive component, and developers must implement, directly in the application code, ways to assign names to cached values, perform lookups, and keep the cache up to date. The extra logic also requires additional testing and debugging time, which can be expensive in some scenarios. The implementation and maintenance of application-level caching are a challenge because developers must refactor all the data access logic to encapsulate cache data into the proper application-level object, and its direct management leads to a concern spread all over the system---mixed with business code---which leads to increasing complexity and maintenance time. Thus, this on-demand and manual caching process can be very time-consuming, error-prone and, consequently, a common source of bugs~\cite{Wang2014}. Gupta et al.~\cite{Gupta2011} and Ports et al.~\cite{Ports2010} both address these implementation issues by providing high-level caching abstractions, in which developers can simply designate application functions as cacheable, and the proposed system automatically caches their results and invalidates the cached data when the underlying source changes.

The second and third issues refer to deciding the right content to cache and the best moment of caching or clearing the cache in order to avoid cache thrashing and stale content. These decisions involve (i) choosing the granularity of cache objects, (ii) translating between raw data and cache objects, and (iii) maintaining cache consistency, which are all tasks to be accomplished by developers and might not be trivial in complex applications. Della Toffola et al.~\cite{DellaToffola2015}, Nguyen and Xu~\cite{Nguyen2013} and Infante~\cite{Infante2014} address part of these issues by identifying and suggesting caching opportunities. However, developers should still review the suggestions and refactor the code, integrating cache logic into the application.

The last issue is related to the maintenance of the cache system, which involves several aspects such as determining replacement policies and the size of the cache. According to Radhakrishnan~\cite{Radhakrishnan2004}, a traditional approach to maintain cache is to set up strategies based on initial assumptions or well-accepted characteristics of workload and access patterns. Thus, cache statistics such as hit ratio, the number of objects cached, and average object size can be collected by observing the application and cache at run-time. These data, compared in the context of desired values of these parameters, can be used to decide whether to change the first choices. Usually, this process is repeated until an acceptable trade-off between configuration and performance is reached, and then cache strategies are fixed for the lifetime of the product or at least for several release cycles. However, its performance may decay over time due to changes in the workload characteristics and access patterns. Therefore, achieving acceptable application performance requires constantly tuning cache decisions, which implies extra time dedicated to maintenance~\cite{Radhakrishnan2004}.

To supply these limitations, recent studies have shown that caching approaches that deal with the changing dynamics of the web environment can perform better than a statically configured cache~\cite{Radhakrishnan2004,Ali2011}. These intelligent approaches try to optimize the cache behavior by dynamically configuring strategies like admission~\cite{Baeza-Yate2007,Surapaneni2011,Rietveld2013,Guerrero2013,Einziger2014}, consistency~\cite{Amza2005} and replacement~\cite{NimrodMegiddo2003,Zhang2003}.

Despite many methods related to intelligent cache approaches have been published~\cite{Podlipnig2003}, adapting cache properties and decisions at application-level has only been tackled by few examples. They provide adaptive solutions for cache update algorithms~\cite{Subramanian2006,Li2006a,Ghandeharizadeh2014,Li2015}, or mapping of requests before repartitioning the cache~\cite{Hu2015}. Some approaches~\cite{Radhakrishnan2004,Qin2013,Wood2013} focus on the infrastructure aspect of application-level caching by exploring the size and fault-tolerance adaptations in cache servers.

Even though these approaches focus on providing an adaptive behavior to application-level cache, none of them takes application specificities into account to autonomously manage their target. They attempt only to optimize cache decisions based on cache statistics like hit-ratio and access patterns at a higher level, thus, ignoring cache meta-data expressed by application-specific characteristics, which are closely related to application computations and could help to reach an optimal performance of the caching service on time. Addressing this issues, content-aware approaches have been proposed~\cite{Pohl2005,Negrao2015} for admitting and replacing items in the cache by exploring descriptive caching hits in the application model or spatial locality (relationships among web objects). Though adaptive caching is in general new and innovative, it is far from being adopted as standard practice in software industry; many improvements must be achieved before this can happen, such as reducing the overhead in terms of resource consumption and processing time of the learning process, and providing easy ways to integrate application and caching method.

Thereby, substantial advances have been made in application-level cache. These design, implementation and maintenance shortcomings call for new methods and techniques, which can support developers to reach a good trade-off between cache development and application performance, leveraging application-specific details. An essential step towards such approaches, taken in this work, is to \emph{understand}, \emph{extract}, \emph{structure} and \emph{document} the application-level caching knowledge implicit and spread in existing web applications. Moreover, our results serve as guidance for the development of application-level caching, given that there is neither off-the-shelf solutions or systematic ways of designing, implementing, and maintaining it.

\section{Study Design}
\label{sec:studydesign}

We performed a qualitative study to understand how application-level caching is implemented in existing software systems that rely on this kind of caching. We provide details of our study in next sections, starting by discussing the study approach in Section~\ref{sec:studyapproach}. We present our goal and research questions in Section~\ref{sec:goalandquestions}, and then describe the study procedure in Section~\ref{sec:procedure}. We introduce the target systems of our study in Section~\ref{sec:targetsystems}, to later proceed to the analysis and interpretation of our obtained results.

\subsection{Study Approach}
\label{sec:studyapproach}

We chose a qualitative method rather than quantitative in order to take a holistic and comprehensive understanding of caching practices adopted by developers, exploring their inner experiences in application development.

Our study was designed based on comparative and interactive principles of \emph{grounded theory}~\cite{Glaser1992}. The purpose of grounded theory is to construct theory grounded in data by identifying general concepts, develop theoretical explanations that reach beyond the known, and offer new insights into the area of study. The systematic procedures of grounded theory enable qualitative researchers to generate ideas. In turn, these ideas can be later studied and verified through traditional quantitative forms of research.

\subsection{Goal and Research Questions}
\label{sec:goalandquestions}

As stated in the introduction, our primary objective while performing this study is to provide guidance to developers when adopting application-level caching in their applications. This study aims to provide such guidance using an \emph{application-centric approach}, i.e.\ by identifying caching solutions that developers usually apply in their applications, taking into account application details. The study was guided by the framework proposed by Basili et al.~\cite{Basili1986, Basili1994}. The paradigm includes the goal-question-metric (GQM) template, which was adopted to define the goal of the study, the research questions to be answered to achieve the goal and metrics for answering these questions.

Although the GQM approach has been extensively used as a way to structure and design empirical studies, it is focused primarily on \emph{quantitative} studies. Given that our study is qualitative rather than quantitative, instead of metrics we specify an evaluation approach with sub-questions, which are used in a similar way as metrics, in the sense of being a way to guide the analysis and answer the mapped questions. This helped us to select the means of achieving our study goal. The description of the study, following the GQM template, is presented in Table~\ref{tab:gqm}.

\begin{table}
    \centering
    \renewcommand\tabcolsep{1mm}
    \caption{Goal Definition (GQM Template).}
    \label{tab:gqm}
    \begin{tabular}{|p{2.7cm}|p{5.5cm}|}
    \hline
    \textbf{Definition Element} & \textbf{Our Study Goal} \\ \hline
    Motivation                  & To identify patterns of application-level caching adopted by developers and understand what kinds of caching implementations and decisions can be automatically inferred, \\ \cline{1-1}
    Purpose                     & characterize and evaluate \\ \cline{1-1}
    Object                      & application-level caching-related design and implementation \\ \cline{1-1}
    Perspective                 & from a perspective of the researcher \\ \cline{1-1}
    Domain: web-based applications     & as they are implemented in the source code and described in issues of web-based applications \\ \cline{1-1}
    Scope                       & in the context of 10 software projects, obtained from open-source repositories and software companies. \\ \hline
\end{tabular}
\end{table}

To achieve our goal, we investigated different application-level caching concerns, which are associated with the three key research questions presented below.

\begin{itemize}
\item[\emph{\textbf{RQ1.}}] What and when is data cached at the application level?
\item[\emph{\textbf{RQ2.}}] How is application-level caching implemented?
\item[\emph{\textbf{RQ3.}}] Which design choices were made to maintain the application-level cache efficient?
\end{itemize}

Determining the cacheable content and the right moment of caching or clearing the cache content are a developer's responsibility and might not be trivial in complex applications, motivating RQ1. In this research question, we aim to identify \emph{what} data is selected to be cached, and the criteria used to detect such cacheable data. Furthermore, this question also explores \emph{when} data should be evicted, as well as constraints, consistency conditions, and the rationale for all these choices.

In addition to design issues, as shown in Figure~\ref{fig:applicationcaching}, the cache system and the underlying source of data are not aware of each other, and the application must implement ways to interact with the cache. Therefore, our goal with RQ2 is to characterize patterns of \emph{how} this implementation occurs in the application code; for example, ways to assigning names to cached values, performing lookups, and keeping the cache up to date.

Finally, determining maintenance strategies to manage efficiently \emph{where} data is placed, such as replacement policies or size of the cache, requires additional knowledge and reasoning from application developers. Nevertheless, there are no foundations to make this choice adequately. Therefore, RQ3 complements the analysis above, questioning how application specificities are leveraged to provide the desired performance to the cache system.

By answering these three central questions, we can extract patterns and guidelines for caching design, implementation and maintenance, and also the context in which each pattern can be adopted.

\subsection{Procedure}
\label{sec:procedure}

As previously described, this study is mainly based on development information of web applications. To investigate different caching constructs, we followed a set of steps to perform our qualitative study: (i) selection of a set of suitable web applications; (ii) specification of a set of questions to guide us in the data analysis and (iii) analysis of each web application using the specified questions. The collected data consists of six different sources of information, explained as follows.

\begin{LaTeXdescription}

\item[Information about the application.] Our goal is to identify caching patterns or decisions, which possibly depend on the application domain. Therefore, we collected general details of the applications to characterize them. The collected application data is (i) its description; (iii) programming languages and technologies involved; and (iii) size of the application in terms of the number of lines of code.

\item[Source code.] Application source code is our core source of information. Since we focus on application-level caching, our analysis is concentrated in the core of the application (i.e.\ the business logic), which is where the caching logic is typically implemented.

\item[Code comments.] Given that caching is an orthogonal concern in the application, unrelated to the business logic, but interleaved with its code, code comments are often used to describe the reasons behind caching implementation and decisions.

\item[Issues.] An issue can represent a software bug, a project task, a help-desk ticket, a leave request form, or even user messages about the project in general. Usually, changes in the code are due to registered issues. Thus, implementation and design decisions are better explained by associated issues in issue platforms, such as GitHub Issue Tracker, JIRA, and Bugzilla.

\item[Developer documentation.] In general, developer documentation consists of useful resources, guides and reference material for developers to learn how the application was implemented. Thus, the available documentation about the project was also collected.

\item[Developers.] In addition to the manual inspection of the above data, we asked developers to which we had access about caching-related implementation, decisions, challenges, and problems.

\end{LaTeXdescription}

To guide the extraction of the information needed from the above data for our qualitative analysis, we derived sub-questions for each research question, which convey characteristics that should be observed and evaluated while looking for answers to the main research questions, i.e.\ sub-questions were used to extract data. Results presented in Section~\ref{sec:analysis} are derived from answers to those questions. Table~\ref{tab:subquestions} depicts the sub-questions derived, which serve as a checklist while analyzing our data.

\begin{table*}
    \centering
    \renewcommand\tabcolsep{1mm}
    \caption{Evaluation Approach (Sub-questions).}
    \label{tab:subquestions}
    \begin{tabular}{|M{.5cm}|p{9cm}|M{2.5cm}|M{2.5cm}|M{2.5cm}|}
        \hline
        \textbf{\#} & \textbf{Sub-question} & \textbf{RQ1. What and when is data cached at the application level?} & \textbf{RQ2. How is application-level caching implemented?} & \textbf{RQ3. Which design choices were made to maintain the application-level cache efficient?} \\ \hline
	1 & What are the motivations to employ cache? & X & & \\ \hline
	2 & What are the typical use scenarios? & X & & \\ \hline
	3 & Where and when data is cached? & X & & \\ \hline
	4 & What are the constraints related to data cached? & X & & \\ \hline
	5 & What are the selection criteria adopted to detect cacheable content? & X & & \\ \hline
	6 & Do developers adopt a pattern to decide which content should be cached? & X & & \\ \hline
	7 & What kinds of bottlenecks are addressed by developers? & X & & \\ \hline
	8 & What motivated the need for explicit caching manipulation? & X & & \\ \hline
	9 & What is the granularity of the cached objects? & X & & \\ \hline
	10 & What is the importance of the cached data to the application? & X & & \\ \hline
	11 & Is there a relationship between the data cached and the application domain? & X & & \\ \hline
	12 & How much memory does the cached data consume? & X & & \\ \hline
	13 & What data is most frequently accessed? & X & & \\ \hline
	14 & How often is the cached data going to be used and changed? & X & & \\ \hline
	15 & What data is expensive? & X & & \\ \hline
	16 & What data depends on user sessions? & X & & \\ \hline
	17 & How up to date does the data need to be? & X & X & \\ \hline
	18 & How is consistency assurance implemented? Why was it chosen? & X & X & \\ \hline
	19 & Where and when is consistency assurance employed? & X & X & \\ \hline
	20 & Which kind of operation/behavior affects cache consistency? & X & X & \\ \hline
	21 & Do developers employ any technique to ease caching implementation, such as design patterns, third-party libraries or aspects? & & X & \\ \hline
	22 & How is the caching logic mixed with the application code? & X & X & \\ \hline
	23 & Is this extra cache logic tested? & & X & \\ \hline
	24 & What is the required format to cache? & & X & X \\ \hline
	25 & How are objects translated to the cache? & & X & X \\ \hline
	26 & How are names (keys) defined for cached objects? & X & X & \\ \hline
	27 & Do developers use another caching layer besides application-level? & & X & \\ \hline
	28 & Is any transparent or automatic caching component being used? & & X & X \\ \hline
	29 & Do developers rely on automatic caching components? & & X & X \\ \hline
	30 & Is it necessary to explicitly manipulate a caching component that should be automatically managed? & & X & X \\ \hline
	31 & Which application layers are more likely to have caching logic? & X & X & \\ \hline
	32 & Do developers perform analysis to measure cache efficiency? & & & X \\ \hline
	33 & What is the replacement policy adopted? Why was it chosen? & & & X \\ \hline
	34 & What is the size of the cache? Why was it defined? & & & X \\ \hline
	35 & What are the default parameter values? & & & X \\ \hline
	36 & Do developers use configurations different from to default? & & X & X \\ \hline
	37 & Do developers take into account application-specific information when defining maintenance strategies? & & X & X \\ \hline
    \end{tabular}
\end{table*}

We followed the analytical process of \emph{coding} in our analysis~\cite{Glaser1992}, which makes it easier to search the data and identify patterns. This process combines the data for themes, ideas and categories, labeling similar passages of text with a code label, allowing them to be easily retrieved at a later stage for further comparison and analysis. We used what we learned from the analysis to adapt our evaluation approach and observation protocols. Insights we had while coding the data and clustering the codes were captured in memos. There are three coding phases in classical grounded theory: open coding, selective coding, and theoretical coding. Open coding generates concepts from the data that will become the building blocks for the theory~\cite{Glaser1992}. The process of doing grounded theory is based on a concept indicator model of constant comparisons of incidents or indicators to incidents~\cite{Glaser1967}. Indicators are actual data, such as behavioral actions and events observed or described in documents and interviews. In this case, an indicator may be an architectural style or design pattern adopted to implement the cache, a data structure, a class, a control flow logic, a comment, a discussion in the issues platform, a paragraph in the documentation, or any other evidence we can get from the data being analyzed.

By using grounded theory, we aimed to construct a well-integrated set of hypotheses that explain how the concepts operated. Thus, the selective coding involves identifying the core category that best explains how study data refers to a large portion of the variation in a pattern and is considered the primary concern or problem related to the study, integrating closely related concepts. Finally, theoretical codes conceptualize how the codes may relate to each other as hypotheses to be incorporated into the theory~\cite{Glaser1992}.

\begin{figure}
    \centering
    \includegraphics[width=\linewidth]{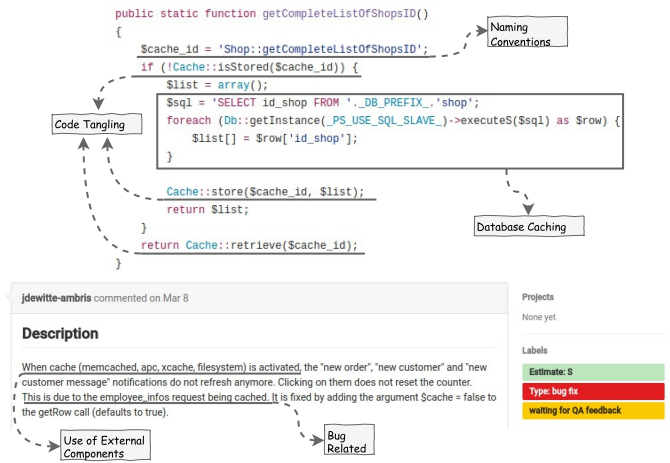}
    \caption{Example of the Analytical Process of \emph{Coding}.}
    \label{fig:codingprocess}
\end{figure}

Figure~\ref{fig:codingprocess} provides an example of such analytical process and illustrates how we collected the different evidences. First, we assigned concepts to pieces of extracted text (open coding), each representing application-level caching characteristics. Figure~\ref{fig:codingprocess} exemplifies different codes identified during the analysis of an application. For instance, \emph{Code Tangling} is created from the observation of cache-related implementation spread all over the application base code (underlined). Then, for each new concept, we verified whether they are connected somehow with existing ones, in order to generate categories (selective coding). Thus, the name assigned to a particular category aims at representing, at a higher abstraction level, all concepts related to it. Regarding the \emph{Code Tangling} example, if \emph{Code Scattered} were found afterwards, we could establish a relationship between the former and the latter to create a category, given that both are related to lack of separation of concerns.

Following the introduced phases of data analysis, we performed mainly a subjective analysis of the data, collecting: (i) typical caching design, implementation and maintenance strategies; (ii) motivations, challenges and problems behind caching, and (iii) characteristics of caching decisions. These collected data were evaluated to conceptualize how the open codes were related to each other as a set of hypotheses in accounting for resolving the primary concern. Furthermore, we also made a broad analysis of the target systems in order to investigate how application-level caching was conceived in them. All the research phases were performed manually, as the collected data (most of them expressed in natural language) analysis is associated with the interpretation of caching approaches.

\subsection{Target Systems}
\label{sec:targetsystems}

In order to investigate different aspects of caching, it was important to select representative systems that make extensive use of application-level caching. To obtain applications that employ application-level caching, we searched through open-source repositories, from which information can be easily retrieved for our study. Based on text search mechanisms, we assessed how many occurrences of cache implementation and issues were present in applications to ensure they would contribute to the study. The more caching-related implementation and issues, the better, so that the rationale behind choices regarding application-level caching could be extracted. Moreover, we managed to obtain commercial applications with partner companies interested in the results of this study.

Therefore, we analyzed systems with a broad range of characteristics. Aiming at reducing the influence of particular software features on the results, we selected systems of different sizes (from 21 KLOC to 250 KLOC, without comments and blank lines), written in different programming languages, adopting different frameworks and architectural styles, and spanning different domains. We studied ten systems in total, of which nine are open-source software projects, and one is from the industry. Due to intellectual property constraints, we will refer to the commercial system as S1. Table~\ref{tab:targetsystems} summarizes the general characteristics of each target system.

\begin{table}
    \renewcommand\tabcolsep{1mm}
    \caption{Target Systems of our Study.}
    \label{tab:targetsystems}
    \centering
    \begin{tabular}{|p{1.8cm}|p{2.7cm}|p{1.3cm}|p{0.9cm}|p{1.1cm}|}
    \hline
        \textbf{Project Name} & \textbf{Domain} & \textbf{Language} & \textbf{KLOC} & \textbf{Analysis Order} \\ \hline
        S1 & Market trend analysis & Ruby & 21 & \#8 \\ \hline
        Pencilblue & CMS and blogging platform & JavaScript & 33 & \#2 \\ \hline
        Spree & e-Commerce & Ruby & 50 & \#7 \\ \hline
        Shopizer & e-Commerce & Java & 57 & \#3 \\ \hline
        Discourse & Platform for community discussion & Ruby & 88 & \#5 \\ \hline
        OpenCart & e-Commerce & PHP & 123 & \#4 \\ \hline
        OpenMRS API and web application & Patient-based medical record system & Java & 127 & \#6 \\ \hline
        ownCloud core & File sharing solution for online collaboration and storage & PHP & 193 & \#10 \\ \hline
        PrestaShop & e-Commerce & PHP & 245 & \#1 \\ \hline
        Open edX & Online courses platform & Python & 250 & \#9 \\ \hline
    \end{tabular}
\end{table}

The open-source applications were selected from GitHub, the widely known common code hosting service. We selected GitHub projects that match the following criteria: (i) projects with some popularity (at least 350 stars); (ii) projects containing application-level caching implementation and issues (at least 50 occurrences of cache-related aspects); (iii) projects written in different programming languages; and (iv) projects of different domains. The first criterion indicates that projects are interesting and were possibly evolved by people other than the initial developers. The second ensures that selected projects would present caching-related implementation and issues, which would contribute more to the study. The inclusion of applications was not restricted to any particular technology or programming language given that the study is focused on identifying language-independent caching patterns and guidelines expressed in the source code. Furthermore, we found applications that use cache in other architectural (e.g.\ database systems) or infrastructure (e.g.\ proxy caching) levels. However, these caching approaches are not handled directly by the developers within the application; consequently, such applications do not fit the purposes of our study. The commercial system was selected by convenience. We selected the applications that better satisfied criteria (i) and (ii), without repeating programming languages and domains. As result, we achieved a set of \emph{representative} systems, required for qualitative studies. In order to minimize bias, applications were analyzed according to a randomly generated order, which is presented in the last column of Table~\ref{tab:targetsystems}.

It is important to highlight that our goal is to capture the \emph{state-of-practice} of application-level caching, by characterizing and deriving caching patterns. Consequently, we assume that the investigated applications---in which caching was introduced in previous releases and was already debugged---have adequate application-level implementations, i.e.\ we did not evaluate the quality of any caching implementation.

\section{Analysis and Results}
\label{sec:analysis}

\newcommand{\uncertaindesignevidence}{(\emph{Evidence 1})\xspace}
\newcommand{\timesensitivecriterionevidence}{(\emph{Evidence 2})\xspace}
\newcommand{\datachangeabilitycriterionvidence}{(\emph{Evidence 3})\xspace}
\newcommand{\userdatacriterionevidence}{(\emph{Evidence 4})\xspace}
\newcommand{\expensivedatacriterionevidence}{(\emph{Evidence 5})\xspace}
\newcommand{\sizeofdatacriterionevidence}{(\emph{Evidence 6})\xspace}
\newcommand{\sizeofcachecriterionevidence}{(\emph{Evidence 7})\xspace}
\newcommand{\smallimprovementsevidence}{(\emph{Evidence 8})\xspace}
\newcommand{\thereisnorightplaceevidence}{(\emph{Evidence 9})\xspace}
\newcommand{\cachinglayerscriterionevidence}{(\emph{Evidence 10})\xspace}
\newcommand{\cachinginboundariesevidence}{(\emph{Evidence 11})\xspace}
\newcommand{\ensuringconsistencyevidence}{(\emph{Evidence 12})\xspace}
\newcommand{\complexdesignevidence}{(\emph{Evidence 13})\xspace}
\newcommand{\simplersolutionsevidence}{(\emph{Evidence 14})\xspace}
\newcommand{\maintenanceproblemevidence}{(\emph{Evidence 15})\xspace}
\newcommand{\bugsevidence}{(\emph{Evidence 16})\xspace}
\newcommand{\codereuseevidence}{(\emph{Evidence 17})\xspace}
\newcommand{\complexnamingevidence}{(\emph{Evidence 18})\xspace}
\newcommand{\namingconventionevidence}{(\emph{Evidence 19})\xspace}
\newcommand{\compleximplementationevidence}{(\emph{Evidence 20})\xspace}
\newcommand{\backgroundprocessevidence}{(\emph{Evidence 21})\xspace}
\newcommand{\externaldependecyevidence}{(\emph{Evidence 22})\xspace}
\newcommand{\additionalcodeevidence}{(\emph{Evidence 23})\xspace}
\newcommand{\externalcomponentbenefitsevidence}{(\emph{Evidence 24})\xspace}
\newcommand{\uncertaintyexternalcomponentevidence}{(\emph{Evidence 25})\xspace}
\newcommand{\sizedefinitionevidence}{(\emph{Evidence 26})\xspace}
\newcommand{\evictiondefinitionevidence}{(\emph{Evidence 27})\xspace}
\newcommand{\tuningperformanceevidence}{(\emph{Evidence 28})\xspace}
\newcommand{\adhoctuningperformanceevidence}{(\emph{Evidence 29})\xspace}

This section details the results of our study and their analysis, according to the research questions we aim to answer. Our collected data consist mainly of source code and issues (expressed in natural language) and, as these are qualitative data, we have undertaken a subjective analysis of the application-level caching (hereafter referred to simply as ``caching'') aspects represented in the target systems. Note that we labeled some findings with ``\emph{Evidence X},'' so that we can later refer to them to support the guidelines we derived from this analysis.

Before addressing each of our research questions, we show an objective analysis of the impact of caching in the investigated applications by identifying all code and issues related to them. This gives a broad sense of how caching is implemented in target systems.

\subsection{Caching in Target Systems}
\label{subsec:cachingintargetsystems}

To investigate how caching is present in target systems, we examined the number of files in which caching logic is implemented, the number of LOC associated specifically with caching (without comments and blank lines), and the number of issues related to it. This analysis is shown in Table~\ref{tab:targetsystemsinfo}, in the columns \emph{\#Cache Files}, \emph{Cache LOC} and \emph{\#Cache Issues}, respectively. This table also gives an overview of further information of each application to which we had access. They are some form of documentation and access to developers. It is important to note that available documentation about caching, in most cases, was limited to an abstract or general description of how caching was adopted. This documentation was available in some open source systems, and we only had access to developers of our system from the industry.

\begin{table}
\centering
\caption{Characteristics of Target Systems.}
\label{tab:targetsystemsinfo}
\begin{tabular}{|p{1.8cm}|p{1cm}|p{1cm}|p{1cm}|p{0.6cm}|p{0.6cm}|}
\hline
\textbf{Project Name} & \textbf{\#Cache Files} & \textbf{Cache LOC} & \textbf{\#Cache Issues} & \textbf{Doc.} & \textbf{Dev.} \\ \hline
S1 & 30 (4.72\%) & 446 (2.12\%) & NA & No & Yes \\ \hline
Pencilblue & 16 (7.04\%) & 526 (1.59\%) & 26 (4.99\%) & No & No \\ \hline
Spree & 27 (2.58\%) & 422 (0.84\%) & 205 (2.92\%) & Yes & No \\ \hline
Shopizer & 31 (4.20\%) & 1725 (3.02\%) & 0 (0\%) & No & No \\ \hline
Discourse & 53 (3.10\%) & 1190 (1.35\%) & 48 (1.34\%) & Yes & No \\ \hline
OpenCart & 44 (4.23\%) & 462 (0.37\%) & 77 (1.97\%) & Yes & No \\ \hline
OpenMRS API and web application & 22 (2.06\%) & 350 (0.27\%) & 19 (1.11\%) & Yes & No \\ \hline
ownCloud core & 240 (10.52\%) & 4344 (2.24\%) & 670 (2.99\%) & Yes & No \\ \hline
PrestaShop & 54 (4.78\%) & 2727 (1.11\%) & 95 (1.96\%) & Yes & No \\ \hline
Open edX & 198 (10.76\%) & 4693 (1.87\%) & 333 (2.95\%) & No & No \\ \hline
\end{tabular}
\end{table}

Based on data presented in Table~\ref{tab:targetsystemsinfo}, we can observe a significant amount of lines of code dedicated to implementing caching, ranging from 0.27\% to 3.02\%. It shows the importance of the caching in the project, considering that caching is a solution for a non-functional requirement (i.e.\ scalability and performance). Furthermore, caching logic is presented in a substantial amount of files, from 2.06\% to 10.76\%, which indicates the caching nature of being spread all over the application.

Moreover, we observed that all analyzed web applications did not adopt caching since their conception. As they increased in size, usage and performance analysis were performed and led to requests for improvements. Thus, developers had to refactor data access logic to encapsulate cache data into the proper application-level object, which is a task that can be very time-consuming, error-prone and, consequently, a common source of bugs. As result, we found a significant number of issues specifically related to caching, achieving the maximum of 4.99\% of the \emph{Pencilblue} issues, which can express in numbers the impact of caching in the entire project \maintenanceproblemevidence.

\begin{figure}
\centering
\includegraphics[width=\linewidth]{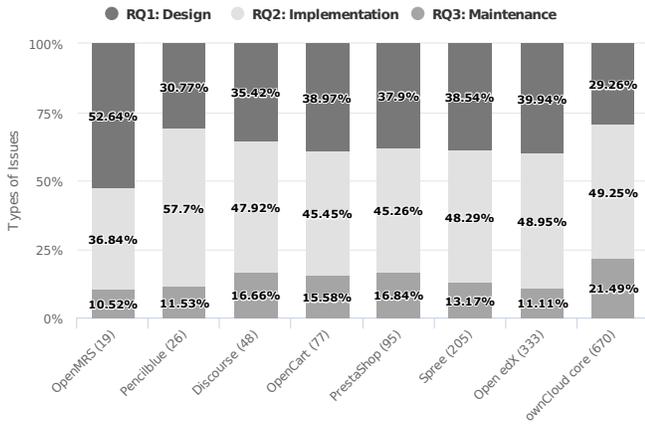}
\caption{Classification of Caching Issues by Topic.}
\label{fig:issuetypes}
\end{figure}

We performed an analysis of the issues available in issue platforms to investigate the primary sources of cache-related problems, typically bugs, in the applications. Based on user messages, code reviews and commit messages that are described in the issues, we classified them into three different categories, which follow the main topic of our research questions: \emph{Design} (e.g.\ changes in the selection of content to be put in the cache), \emph{Implementation} (e.g.\ bugs in the implementation) and \emph{Maintenance} (e.g.\ performance tests, adjustments in replacement algorithms or size limits of the cache). Results of the performed analysis are presented in Figure~\ref{fig:issuetypes}, in which applications are ordered ascending by the number of cache-related issues, which is shown next to each application name. Only open source projects with issues related to caching are detailed in this figure.

As can be seen in Figure~\ref{fig:issuetypes}, caching implementation and associated design decisions are much more discussed and revised by developers than maintenance decisions. Caching implementation, which is spread in the code and involves the choice for appropriate locations to add and remove elements from the cache, is error-prone and can compromise code legibility. Consequently, many issues are associated with bug fixes, technological details and code refactorings. Moreover, despite being less frequent, caching design is time-consuming and challenging, given that it requires understanding of the application behavior, as well as limitations, conditions and restrictions of content being cached. In applications analyzed, the mean (M) and standard deviation (SD) values of cache-related implementation issues are M = 47.45\% and SD = 5.76\%, while design issues achieve M = 37.92\% and SD = 7.11\%. Finally, because fine-grained configurations require empirical analysis such as cache profiling, and there is little evidence that this was performed in investigated applications, maintenance decisions often result in the choice for default settings. Consequently, a lower number of issues is associated with such decisions, specifically M = 14.61\% and SD = 3.44\%. This issue analysis allowed us to understand the aspects of caching that require more effort from developers.

After this broad analysis of our target systems, we further analyze them focusing on our research questions. For each research question, we detail coding results and explore the answers achieved while analyzing emerged codes.

\subsection{RQ1: What and when is data cached at the application level?}
\label{subsec:RQ1}

In this question, we focus on going beyond the facts exposed by source code and analyze the reasoning behind caching decisions such as what and why data is cached or evicted, when and where caching is done, and what are the conditions and constraints involved with this. Therefore, issues, code comments, and documentation about the cache of applications were the primary source of information to find answers to this question, because they convey (in natural language) the rationale behind implementation decisions.

With the analysis of all provided information about application-level caching, we gathered and categorized it in an exhaustive way, observing fine-grained details. As said, we followed the principles of grounded theory, and its first phase consists of open coding. As a result, we identified 72 initial categories. Following the process, these categories were not predefined and were created and reviewed during the open coding process.

\begin{figure}
\centering
\includegraphics[width=\linewidth]{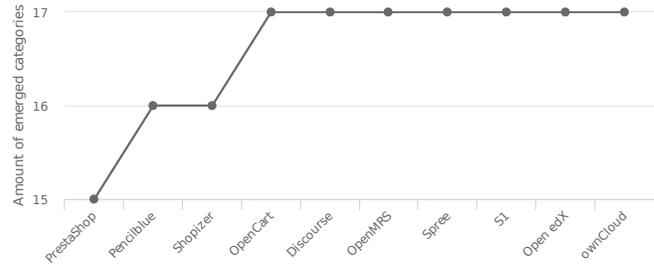}
\caption{Amount of Emerged Categories based on the Analysis of Each System.}
\label{fig:plateau}
\end{figure}

These initial categories that emerged during open coding were posteriorly analyzed and conceptualized, focusing on identifying relationships between them as hypotheses and grouping categories that may be theoretically coded as causal and associated with degrees. The open codes were directly used to derive more abstract categories during the theoretical coding. In the end, we identified 17 concepts, which better convey the implicit knowledge about application-level caching design, implementation, and maintenance acquired from the applications. After the analysis of the fourth application, codes stabilized, as shown in Figure~\ref{fig:plateau}.

\begin{figure*}%
\centering%
\subfloat[Coding Occurrences.]{%
    \includegraphics[width=\textwidth]{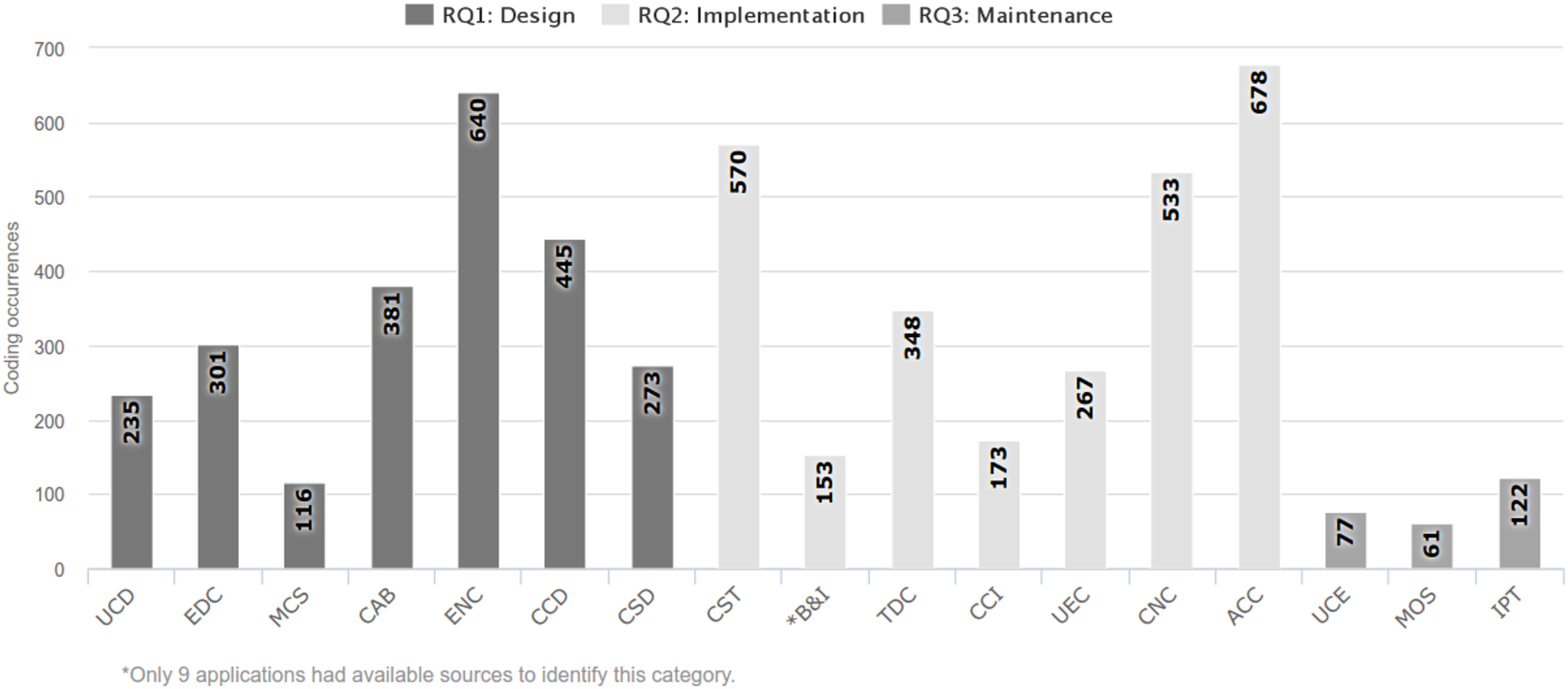}%
    \label{fig:codingamount}%
}%
\hfil%
\subfloat[Coding Percentage by Application.]{%
    \includegraphics[width=\textwidth]{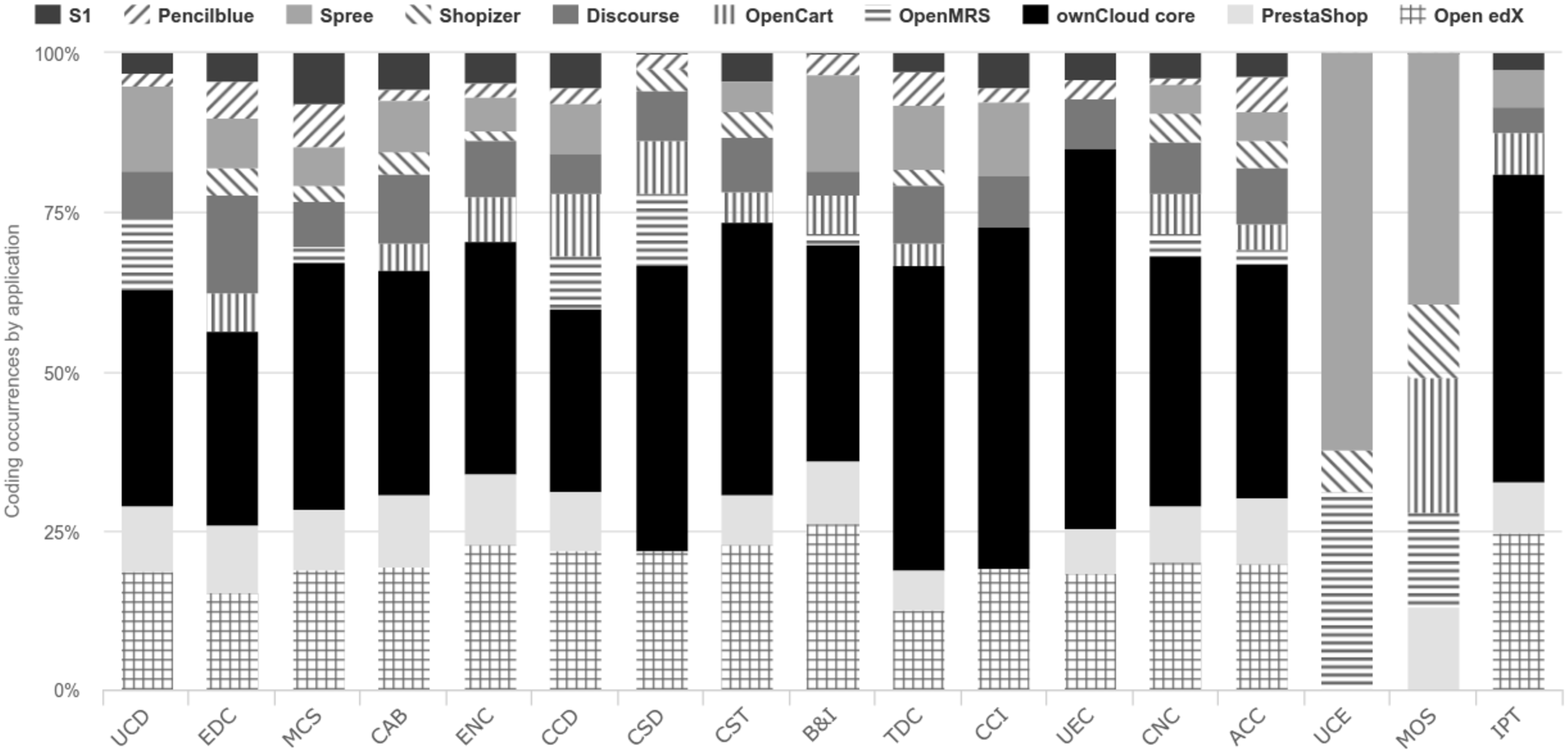}%
    \label{fig:codingpercentage}%
}%
\caption{Categories Emerged from the Study.}%
\label{fig:coding}%
\end{figure*}

The theoretical categories are described in Table~\ref{tab:codingRQ1}, which presents their description, an original piece of content (example of evidence) and sources of data classified in each category. Moreover, categories are also shown in Figure~\ref{fig:codingamount} with the associated number of occurrences in the applications analyzed and classified according to each research question. Figure~\ref{fig:codingpercentage} shows the percentage of contribution of each application to the categories emerged from the study. These figures also present categories identified in \emph{RQ2} (described in Table~\ref{tab:codingRQ2}) and \emph{RQ3} (described in Table~\ref{tab:codingRQ3}), which are discussed in the following sections. Acronyms used in Figures~\ref{fig:codingamount} and~\ref{fig:codingpercentage} are introduced in Tables~\ref{tab:codingRQ1}, \ref{tab:codingRQ2}, and~\ref{tab:codingRQ3}.

\begin{table*}
\centering
\renewcommand\tabcolsep{1mm}
\caption{Analysis of RQ1: Emerged Design Categories.}
\label{tab:codingRQ1}
\begin{tabular}{|p{2.2cm}|p{5.3cm}|p{6cm}|c|c|c|c|c|}
\hline
\textbf{Category} & \textbf{Description} & \textbf{Example of Evidence} & \multicolumn{5}{c|}{\textbf{Evidence Sources}} \\ \cline{4-8}
\textbf{(Acronym)} &                      &                              & \textbf{SC} & \textbf{COM} & \textbf{IS} & \textbf{DOC} & \textbf{DEV} \\ \hline
                  
Uncertainty in{\par}Cache Design (UCD) & Indication, by developers, of uncertainty regarding cacheable content. & A code comment before an expensive operation: \par \emph{TODO cache the global properties to speed this up??} & & X & X & & X \\ \hline

Explanations of{\par}Cache Design{\par}Choices (EDC) & Explanations provided by developers regarding cache design choices, such as specific criteria to determine cacheable content. & A code comment detailing the motivation for caching: \par \emph{Fetch all dashboard data. This can be an expensive request when the cached data has expired, and the server must collect the data again.} & X & X & X & X & \\ \hline

Multiple Caching Solutions (MCS) & Occurrences of multiple caching solutions, which can involve different third-party components or different application layers. As a consequence, the same content can be cached at different places in varying forms. & A user announcement on issue platform: \par \emph{Caching has now landed: Fragment caching for each product; Fragment caching for the lists of products in home/index and products/index; Caching in the ProductsController, using expires\_in which caches for 15 minutes.}  & X & X & X & X &   \\ \hline

Caching in{\par}Application Boundaries (CAB) & Occurrences of caching content being added to the cache because they retrieve data from frameworks, libraries, or external applications. & A sentence in the documentation: \par \emph{Most stores spend much time serving up the same pages over and over again. [...] In such cases, a caching solution may be appropriate and can improve server performance by bypassing time-consuming operations such as database access.}  & X & X & X & X & X \\ \hline

Ensuring{\par}Consistency (ENC) & Design of some kind of consistency approach such as expiration policies and invalidation, preventing stale data. & A user message on issue platform detailing the invalidation approach adopted: \par \emph{Ideally we cache until the topic changes (a new post is added, or a post changed) [...] less ideally we cache for N minutes} & X & X & X & X & X \\ \hline

Complex Cache{\par}Design (CCD) & Indication that a cache design choice is difficult to understand, confusing developers and requiring detailed comments. & A user message on issue platform: \par \emph{i can see that it may speed up the query responses, but is the saved time substantial enough to be worth the effort?} & & & X & & X \\ \hline

Choice for{\par}Simple Cache{\par}Design Solutions (CSD) & Indication that developers selected simple solutions for cache rather than complex ones, such as defining a time-to-live (TTL) instead of implementing manual invalidation, possibly in order to balance design effort and caching gains. & A code snippet defining a default setting: \par \emph{\textless defaultCache maxElementsInMemory=``1000'' \newline eternal=``false'' timeToIdleSeconds=``60'' \newline timeToLiveSeconds=``0'' overflowToDisk=``false'' \newline diskPersistent=``false''/\textgreater} & X & X & X & & X  \\ \hline

\multicolumn{8}{l}{\emph{Labels of Evidence Sources}: \textbf{SC}-Source Code (without comments); \textbf{COM}-Code Comments; \textbf{IS}-Issues; \textbf{DOC}-Documentation; \textbf{DEV}-Developers.}
\end{tabular}
\end{table*}

We observed in eight from the ten analyzed applications that developers indicated that they had \emph{uncertainty in cache design}, with respect to deciding what data should be cached and the right moment to do it, causing missed caching opportunities. There are comments in the source code and issues about whether to cache some specific data or not, showing that the connection between observed bottlenecks in the application and opportunities for caching is not straightforward and requires a deeper analysis \uncertaindesignevidence.

Therefore, selection criteria based on common sense or past experiences are initial assumptions to ease such decisions. Despite these criteria are usually unjustified, they guide developers while selecting content. We observed in 90\% of the applications the definition of selection criteria to make the distinction of cacheable from uncacheable content easier. These criteria were observed in \emph{explanations of cache design choices} in comments, issues, and documentation. We identified common criteria used to determine whether to cache or not a specific content, which are: (i) content change frequency \timesensitivecriterionevidence, (ii) content usage frequency \datachangeabilitycriterionvidence, (iii) content shareability \userdatacriterionevidence, (iv) content retrieval complexity \expensivedatacriterionevidence, (v) content size \sizeofdatacriterionevidence, and (vi) size of the cache \sizeofcachecriterionevidence.

Regarding design choices, we observed common practices. The first is associated with the lack of a specific approach to cache data. To process a client request, application components of distinct layers and other systems (databases, web services, and others) are invoked, and each interaction results in data transformation, which is likely cacheable. Due to this, nine analyzed applications present \emph{multiple caching solutions} by not specifying cacheable layers, components or data, and employing cache wherever can potentially provide performance and scalability benefits \smallimprovementsevidence, no matter which is the application layer, component or data \thereisnorightplaceevidence. As a consequence, the same content can be cached at different places, from the database to controllers, in varying forms such as query results, page fragments or lists of objects \cachinglayerscriterionevidence.

The second design choice is associated with the concern of reducing the communication latency between the application and other systems, which increases the overall response time of a user request. Therefore, we noticed \emph{caching in application boundaries} in nine of the analyzed applications, addressing remote server calls and requests to web services, database queries, and loads dependent on file systems, which are common bottlenecks \cachinginboundariesevidence.

Despite choosing where and what to cache, cached values are valid only as long as the sources do not change, and when sources change, a consistency policy should be employed to ensure that the application is not serving stale data. Therefore, in nine analyzed applications, there are indications that developers demand an effort to design \emph{consistency approaches}, reasoning about the lifetime of cached data, as well as eviction conditions and constraints. We identified common approaches to keep consistency, which are: (i) a less efficient and straightforward approach is to invalidate cached values based on mapping actions that change its dependencies and, after a change is invoked, invalidate cached values and recompute them at the next request; and (ii) the use of a less intrusive alternative such as a time-to-live (TTL) or a replacement approach, which require a certain level of staleness \ensuringconsistencyevidence.

All these caching design options may become complex and difficult to understand. Indeed, identifying caching opportunities and ensuring consistency can add much code and may not be trivial to implement and understand. Due to the nature of application-level caching, such logic is spread all over the system. We noticed in 90\% of the applications the presence of pieces of evidence that \emph{caching design achieves a high level of complexity}, requiring many comments explaining even less complex parts and issues related to the lack of understanding of caching decisions \complexdesignevidence.

Due to the increasing demand and complexity involved with caching, developers try to decrease cache-related effort by \emph{adopting simple design solutions}. We observed in seven of the ten analyzed applications design choices such as managing consistency based on expiration time, keeping default parameters, selecting data without any criteria, or even adopting external solutions, which do not claim extra reasoning, time, and modifications to the code \simplersolutionsevidence.

We discussed our findings regarding cache design decisions and we next discuss observations made associated with how to implement design decisions.

\subsection{RQ2: How is application-level caching implemented?}
\label{subsec:RQ2}

To understand how developers integrate application-level caching logic in their applications, we analyzed the cache implementations from two perspectives. The first consists of examining the explicit caching logic present in the application code, focusing on analyzing where the caching logic is placed, for what this logic is responsible, and when it is executed. The second evaluates the integration and communication between the application and an external caching component, which is usually used to relieve the burden on developers, easing cache implementation, raising abstraction levels, or even taking full control of caching. For this question, the most valuable data comes from source code and comments, which express implementation details. The theoretical categories referring to \emph{RQ2} are described in Table~\ref{tab:codingRQ2} and shown in Figure~\ref{fig:coding}.

\begin{table*}
    \centering
     \renewcommand\tabcolsep{1mm}
\caption{Analysis of RQ2: Emerged Implementation Categories.}
\label{tab:codingRQ2}
\begin{tabular}{|p{2cm}|p{4.9cm}|p{6.6cm}|c|c|c|c|c|}
\hline
\textbf{Category} & \textbf{Description} & \textbf{Example of Evidence} & \multicolumn{5}{c|}{\textbf{Evidence Sources}} \\ \cline{4-8}
\textbf{(Acronym)} &                      &                              & \textbf{SC} & \textbf{COM} & \textbf{IS} & \textbf{DOC} & \textbf{DEV} \\ \hline

Code Scattering and Tangling (CST) & Presence of caching logic spread all over the application, indicating a lack of separation of concerns. & A developer quote: \par \emph{At first glance, the cache code hinders the understanding of the business logic. Also, the cache logic itself it is not easy to get.} & X & X & X &   & X \\ \hline

Bugs and Issues (B\&I) & Presence of bugs and problems due to caching reporting, for example, increased client response time, reduced throughput, increased server resource utilization. & A bug report on an issue platform: \par \emph{When you import categories with a parent category which does not exist, it prevents from duplicate it because of the cache.}  &   &   & X &   &   \\ \hline

Technical Debt{\par}Concerns (TDC) & Indication of extra effort spent by developers to build an extensible and well designed caching component. & A user message on an issue platform: \par \emph{my concern with this implementation is that it can lead to some confusing situations. [...] I think the caching needs to be more granular}  & X & X & X & X & X \\ \hline

Complex Caching Implementation (CCI) & Presence of complex constructs such as batch processing and asynchronous communication, which require extra effort and reasoning from developers to be implemented. & A user message on an issue platform: \par \emph{Can you give me some tips on how to safely implement an async step to save this? [...] I could have the end user polling and refreshing against that as needed.} & X & X & X & X &  \\ \hline

Use of External Components (UEC) & Indication of use of third-party caching solutions to help the implementation, raising the level of abstraction of some caching aspects. & A sentence in documentation: \emph{We use Redis [a third-party solution] as a cache and for transient data.}  & X & X & X & X & X \\ \hline

Complex Naming Conventions (CNC) & Choice for complex keys of caching content, causing developers to spend time and effort to elaborate and understand such keys. & A code snippet: \par \emph{cache\_id = 'objectmodel\_' . \$entity\_defs['classname'] . '\_' . (int)\$id . '\_' . (int)\$id\_shop . '\_' . (int)\$id\_lang;}  & X & X & X &   &   \\ \hline

Additional Caching Code (ACC) & Code implemented to support the caching logic, such as implemented caching tests, logic to monitor cache statistics and additional interfaces to support available caching providers. & A code snippet exposing a test of caching logic: \par \emph{it "can set and get false values when return cache nil" do \newline
    @store.set :test, false \newline
    expect(@store.get(:test)).to be false\newline
  end}  & X & X & X &   &   \\ \hline

\multicolumn{8}{l}{\emph{Labels of Evidence Sources}: \textbf{SC}-Source Code (without comments); \textbf{COM}-Code Comments; \textbf{IS}-Issues; \textbf{DOC}-Documentation; \textbf{DEV}-Developers.}
\end{tabular}
\end{table*}

We observed in eight from the ten analyzed applications that they present \emph{code scattering and tangling}, on caching logic, causing low cohesion and high coupling in the code. Caching control code, responsible for caching data in particular places, was spread all over the base code, being invoked when application requests were processed. Consequently, there is a lack of separation of concerns, leading to increased maintenance effort \maintenanceproblemevidence.

A possible cause for this is that caching was not part of all the applications since their conception. Applications were developed and, as they evolved, usage and performance problem reports led to requests for response time improvements. Thus, developers had to refactor existing business logic to include caching aspects. 
Interleaved caching logic can cause not only increased maintenance effort but also be a source of bugs---in eight (from nine, given that we had no access to an issue tracker of one of the applications) analyzed applications, there are issues associated with \emph{bugs due to caching} \bugsevidence.

Although this problem is present in the code, there are indications that developers know about it and express they are willing to improve the provided solution. In order to reduce the impact of an infrastructure component to the system business logic, we identified cases where there are suggestions to design more extensible classes and modules, refactoring and reducing cache-related code, and reusing components \codereuseevidence. This acknowledgment of \emph{technical debt} was observed in 90\% of the applications.

Regarding implementation choices, we observed common practices. The first is associated with how to name cached data. In order to use in-memory caching solutions, there is no prescribed way to organize data. Typically, unique names are assigned to each cached content, thus leading to a key-value model---and this was the case in all investigated applications. Given that cache stores lots of data, the set of possible names must be large; otherwise, two names (keys) can conflict with each other and, thus stale (or even entirely wrong) data can be retrieved from the cache. Consequently, \emph{complex naming conventions} were adopted \complexnamingevidence.

Therefore, in all applications, there are indications, in the form of source code, comments, and issues, that developers demand an effort to understand and improve cache keys, reasoning about alternative better ways to organize and identify content. We identified common content properties used to build cache keys, which are: (i) content \emph{ids}, (ii) method signatures, and (iii) a tag-based identification, in which the type of content, the class, the module or hierarchy are used \namingconventionevidence.

The second implementation choice is associated with the concern of not increasing the throughput due to the introduction of communication between the cache and the application. Due to this, six of our applications make \emph{use of non-trivial programming techniques} to ensure a high performance caching implementation, taking advantage of every improvement opportunity \compleximplementationevidence. Solutions related to this include processing caching requests in the background, batching caching calls together, or even implementing a coarse-grained caching service that allows a single logical operation to be performed by using a single round trip. These prevent blocking client requests when a possibly large caching processing is being performed, e.g.\ a caching warm up or update \backgroundprocessevidence.

Given that implementing cache is challenging, we noticed that in six applications developers made \emph{use of supporting libraries and frameworks}. This was done to prevent adding much cache-related code to the base code, because such components raise the abstraction level of caching, providing some ready-to-use features \externaldependecyevidence. Examples of such external components are distributed cache systems, e.g.\ Redis~\cite{Redis2016} and Memcached~\cite{Memcached2016}, and libraries that can act locally, e.g.\ Spring Caching~\cite{SpringCaching2016}, EhCache~\cite{EhCache2016}, Infinispan~\cite{Infinispan2016}, Rails low-level caching~\cite{RailsCaching2016}, Google Guava~\cite{Guava2016} and Caffeine~\cite{Caffeine2016}.

Such supporting libraries and frameworks not only provide partial ready-to-use features but also reduce the amount of additional effort required to guarantee that the cache is working. We observed in all applications that the cache includes \emph{code dedicated to test, debug and configure cache components}, which can be expensive in some scenarios \additionalcodeevidence.

We next discuss observations made associated with how the cache is maintained and tuned.

\subsection{RQ3: Which design choices were made to maintain the application-level cache efficient?}
\label{subsec:RQ3}

After designing and implementing the cache, maintenance issues are still open. Initial assumptions concerning caching design can become invalid as changes occur in the application domain, workload characteristics or access patterns, thus leading to a performance decay. Therefore, in order to keep the cache efficient, it is necessary to constantly tune cache settings, at least in between releases. Maintenance decisions involve: (i) detecting improvement opportunities in the cache design, (ii) dealing with specificities of the cache deployment scheme (e.g. local or remote, shared or dedicated, in memory or on disk storage), (iii) determining the appropriate size of the cache, (iv) defining how many objects are allowed to be cached at same time, and (v) what should be done in case of the cache is full. The theoretical categories identified while investigating how developers approach these issues and maintain cache efficient are described in Table \ref{tab:codingRQ3} and shown in Figure~\ref{fig:coding}.

\begin{table*}
    \centering
     \renewcommand\tabcolsep{1mm}
\caption{Analysis of RQ3: Emerged Maintenance Categories.}
\label{tab:codingRQ3}
\begin{tabular}{|p{1.9cm}|p{5.6cm}|p{6cm}|c|c|c|c|c|}
\hline
\textbf{Category} & \textbf{Description} & \textbf{Example of Evidence} & \multicolumn{5}{c|}{\textbf{Evidence Sources}} \\ \cline{4-8}
\textbf{(Acronym)} &                      &                              & \textbf{SC} & \textbf{COM} & \textbf{IS} & \textbf{DOC} & \textbf{DEV} \\ \hline

Uncertainty in \par External Cache Component (UCE) & Indication of uncertainty in the behavior of an external cache component, when the applications rely on it instead of handling the cache manually. & A code comment: \par \emph{Is hibernate taking care of caching and not hitting the db every time? (hopefully it is)} & X & X & X & X & \\ \hline

Maintenance \par of Cache Size (MOS) & Indication of choices made in order to control the cache size, such as definitions of eviction approaches to fit cache size limitations. & A user message on issue platform: \par \emph{Be sure the local cache will not grow out of control (especially during big operations like product import)} & X & X & X & & \\ \hline

Improvements based on \par Performance Tests (IPT) & Indication that improvements were made based on performance tests. & An user message on issue platform comparing two request log traces, with and without cache: \par \emph{My results were something like: 1) Using rows: 30ms, 5mb RAM (peak usage); 2) Fetching all at once: 3ms, 5.75MB RAM} & & & X & & X \\ \hline

\multicolumn{8}{l}{\emph{Labels of Evidence Sources}: \textbf{SC}-Source Code (without comments); \textbf{COM}-Code Comments; \textbf{IS}-Issues; \textbf{DOC}-Documentation; \textbf{DEV}-Developers.}
\end{tabular}
\end{table*}

As mentioned in Section \ref{subsec:RQ1}, we observed that developers often rely on external components, which can raise the level of caching abstraction, providing some cache-related components ready-to-use. The most common tasks delegated to external components are associated with cache space allocation and management, since maintaining cache manually through lists or hashes inside the application involves issues such as dealing with concurrency, key management and size limits, which may not be trivial.

Furthermore, in cases where updates in the base code are not an option (due to time or technical restrictions), a transparent and automatic caching component can provide fast results. These solutions address layers before and after application boundaries, and require only a few adaptations to the application needs \externalcomponentbenefitsevidence.

However, we observed in three analyzed applications an \emph{uncertainty regarding the quality, in terms of performance, of a transparent component} \uncertaintyexternalcomponentevidence. In fact, transparent caching solutions are out of the application control and are built and maintained by a third party community. Despite they provide ways to developers to manipulate and observe their behavior, it may become a problem when requests for improvements force developers to customize the behavior of the built-in cache component, thus creating a new and particular behavior for the component, which is supposed to be generic and easy-to-use. All this manipulation should also be tested and evaluated, leading to the generation of test cases to assert the reliability of the external component being used.

The use of transparent and automatic caching solutions relieve developers also from two maintenance tasks. First, while it is true that the larger the cache size, the better the performance, very large caches are unrealistic. While cost is the primary reason for limiting the size of a cache, efficient invalidation approaches play a key role as well. We noticed in five of our analyzed applications indications that developers explicitly \emph{deal with size limitations} \sizedefinitionevidence, by defining a specific size of the cache, replacement policies \evictiondefinitionevidence and the number of manageable objects allowed to be cached.

The second maintenance task refers to cache performance improvements. We observed in seven of the ten applications indications that developers \emph{perform tests}, comparing the application behavior with and without a particular caching approach \tuningperformanceevidence. However, while focusing on tuning all the aspects of caching, developers perform trivial analyses, which are usually based on comparing logs or execution of several requests in a row and measurement of response times. These tests are inappropriately reported and documented, becoming irreproducible. As result, these tests can lead to false-positive improvements, based on poor and biased conclusions \adhoctuningperformanceevidence.

Based on our analysis, we identified application-level caching decisions and behaviors adopted by developers. Our findings allowed us to propose a set of guidelines and patterns for the development of a caching approach, which are described in next section.

\section{Guidelines and Patterns}
\label{sec:guidelinesandpatterns}

This study allowed us to understand how developers deal with application-level caching in their applications, by explaining design, implementation and maintenance choices. Our findings and observations were used as foundation to the provision of practical guidance for developers with respect to caching. In this section, we thus introduce guidelines, which are general rules to be followed by developers while developing application-level caching, and patterns, which are structured reusable solutions to recurring problems in cache design, implementation or maintenance. For each guideline, we indicate evidences that support it. Both guidelines and patterns derived from our study are classified into categories, explained below. Note that there are patterns associated with the proposed guidelines.

\begin{LaTeXdescription}
\item[Design.] Support to design decisions associated with application-level caching.
\item[Implementation.] Support to implementation issues of application-level caching, by providing solutions and guidance at the code level.
\item[Maintenance.] Support to performance analysis and improvement of application-level caching.
\end{LaTeXdescription}

\subsection{Design Guidelines}

\emph{\textbf{Evaluate different abstraction levels to cache.}} \thereisnorightplaceevidence and \cachinginboundariesevidence It is important to cache data where it reduces the most processing power and round trips, choosing locations that support the lifetime needed for the cached items, despite where it is located in the application. Different levels of caching provide different behavior, and possibilities must be analyzed. For instance, caching in the model or database level offers higher hit ratios, while caching in presentation layer can reduce the application processing overhead significantly in the application in case of a hit. However, in the latter case, hit ratios are in general lower. It is possible to cache data at various layers of an application, according to the following layer-by-layer considerations.

\begin{LaTeXdescription}
  \item[Controller layer.] Caching data at the controller layer should be considered when data needs to be frequently displayed to the user and is not cached on a per-user basis. At this level, controllers usually work by serving parameterized content, which can be used as an identifier in the cache. For example, if a list of states is presented to the user, the application can load these once from the database and then cache them, according to the parameters passed in the first request.
  \item[Business or service layer.] Caching data at the business layer should be considered if an application needs to process requests from the presentation layer or when the data cannot be efficiently retrieved from the database or another service. It can be implemented by using hash tables, library or framework. However, at this level, a large amount of data tends to be manipulated and caching it can consume more memory and leads to memory bottlenecks.
  \item[Model or database layer.] At the model or database layer, a large amount of data can be cached, for lengthy periods. It is useful to cache data from a database when it demands a long time to process queries, avoiding unnecessary round-trips.
\end{LaTeXdescription}

\emph{\textbf{Stack caching layers.}} \thereisnorightplaceevidence and \cachinglayerscriterionevidence It is reasonable to say that the more data cached, the lower the chance of being hit without any content already loaded. Caching might be at the client, proxy server, inside the application in presentation, business, and model logics, or database. Despite the same data may be cached in multiple locations, when the cache expires in one of them, the application will not be hit with an entirely uncached content, avoiding processing and network round trips. However, it is important to keep in mind that caching layers imply a range of responsibilities, such as consistency conditions and constraints, and extra code and configuration. Due to this, it is important to consider many caching layers but, at the same time, achieve a good trade-off between caching benefits and implementation effort.

\emph{\textbf{Separate dynamic from static data.}} \datachangeabilitycriterionvidence and \timesensitivecriterionevidence Content can be distinguished in static, dynamic, and user-specific. By partitioning the content, it is easier to select portions of the data to cache.

\emph{\textbf{Evaluate application boundaries.}} \cachinglayerscriterionevidence and \cachinginboundariesevidence Communication between application and external components is a common bottleneck and, consequently, an opportunity for caching. Consider caching for database queries, remote server calls and requests to web services, which are made across a network.

\emph{\textbf{Specify selection criteria.}} \uncertaindesignevidence, \timesensitivecriterionevidence, \datachangeabilitycriterionvidence, \expensivedatacriterionevidence, \sizeofdatacriterionevidence and \sizeofcachecriterionevidence Selecting the right data to cache involves a great reasoning effort given that data manipulated by web applications range in dynamicity, from being completely static to changing constantly. To optimize this selection process, there are four primary selection criteria used by developers while detecting cacheable content, which should be used in decisions regarding whether to cache. These criteria are described below, ordered according to their importance; i.e.\ the higher the influence level, the earlier it is presented.

\begin{LaTeXdescription}
  \item[Data change frequency.] Developers should seek for data that have some degree of stability, i.e.\ those that are more used than changed. Even if data are volatile and change in time intervals, caching still brings a benefit. This is the first factor to be considered since caching volatile data implies the implementation of consistency mechanisms, which is not trivial and requires an extra effort and reasoning from developers. In short, the cost of consistency approaches cannot be higher than the benefit of caching. Besides, when stale data is not a critical issue, an approach of weak consistency can be employed, such as time-to-live (TTL) eviction, where data is expired after a time in cache, regardless of possible changes.

  \item[Data usage frequency.] Frequent requests, operations, queries and shared content (accessed by multiple users) must be identified, focusing on recomputation avoidance. Even if some processing can be fast enough at a glance, it can potentially become a bottleneck when being invoked many times. Despite being frequently used, user-specific data cannot be shared and may not bring the benefit of caching, being usually left out of the cache.
  
  \item[Data retrieval complexity.] Data that is expensive to retrieve, compute, or render, regardless of its dynamicity, is always considered a good caching opportunity.

  \item[Size of the data.] The size of the content being cached should be considered when using size-limited caches. In this case, an adequate trade-off between popularity (hits) and size of the items must be achieved. Keeping small popular items in the cache tends to optimize hit-ratio; however, a hit in a large item may be more beneficial for an application than many hits on small items. At the same time, filling the cache with few large items may turn the cache performance dependent on a good replacement policy.
\end{LaTeXdescription}

\emph{\textbf{Evaluate staleness and lifetime of cached data.}} \timesensitivecriterionevidence, \datachangeabilitycriterionvidence and \ensuringconsistencyevidence Every piece of cached data is already potentially stale, it is important to rethink the degree of integrity and potential staleness that the application can compromise for increased performance and scalability. Many cache implementations adopted an expiration policy to invalidate cached data based on a timeout since weak consistency is easier than defining a hard-to-maintain, but more robust, invalidation process. In short, developers must ensure that the expiration policy matches the pattern of access to applications that use the data, which is based on determining how often the cached information is allowed to be outdated, and relaxing freshness when possible.

\emph{\textbf{Avoid caching per-user data.}} \userdatacriterionevidence and \sizedefinitionevidence It is recommended to avoid caching per-user data unless the user base is small and the total size of the cached data does not require an excessive amount of memory; otherwise, it can cause a memory bottleneck. However, if users tend to be active for a while and then go away again, caching per-user data for short-time periods may be an appropriate approach. For instance, a search engine that caches query results by each user, so that it can page through results efficiently.

\emph{\textbf{Avoid caching volatile data.}} \timesensitivecriterionevidence and \datachangeabilitycriterionvidence Data should be cached when it is frequently used and is not continually changing. Developers should remember that caching is most effective for relatively stable data, or data that is frequently read. Caching volatile data, which is required to be accurate or updated in real time, should be avoided.

\emph{\textbf{Do not discard small improvements.}} \datachangeabilitycriterionvidence, \smallimprovementsevidence and \thereisnorightplaceevidence The user perceived latency is reduced by any caching solution employed. This means that even not obvious scenarios should be target of caching, i.e.\ it is not true that solely data that is frequently used and expensive to retrieve or create should considered for caching. Furthermore, data that is expensive to retrieve and is modified on a periodic basis can still improve performance and scalability when properly managed. Caching data even for a few seconds can make a large difference in high volume sites. If the data is handled more often than it is updated, it is also a candidate for caching.

\subsection{Implementation Guidelines}

\emph{\textbf{Keep the cache API simple.}} \maintenanceproblemevidence, \codereuseevidence, \compleximplementationevidence and \externaldependecyevidence Caching logic tends to be spread all over the application, and a good solution should be employed to avoid writing messy code at the cost of high maintenance efforts.

\emph{\textbf{Define naming conventions.}} \namingconventionevidence and \complexnamingevidence To define appropriate names for cached data, it is important to assign a name that is related to its context, the data itself, and the caching location. It can provide two direct benefits: (a) prevention of key conflicts, and (b) guidance of cache actions such as updates and deletes of stale data in case of changes in the source of information.

\emph{\textbf{Perform cache actions asynchronously.}} \backgroundprocessevidence For large caches, it is adequate to load the cache asynchronously with a separate thread or batch process. Moreover, when an expired cache content is requested, it needs to be repopulated and doing so synchronously affects response time and blocks the request processing thread.

\emph{\textbf{Do not use cache as data storage.}} \evictiondefinitionevidence An application can modify data held in a cache, but the cache should be considered as a transient data store that can disappear at any time. Therefore, developers should not save valuable data only in the cache, but keep the information where it should be as well, minimizing the chances of losing data if the cache unexpectedly becomes unavailable.

\subsection{Maintenance Guidelines}

\emph{\textbf{Perform measurements.}} \uncertaindesignevidence, \smallimprovementsevidence, \sizedefinitionevidence, \tuningperformanceevidence and \adhoctuningperformanceevidence Caching is an optimization technique and, as any optimization, it is important perform measurements before making substantial changes, given that not all application performance and scalability problems can be solved with caching. Furthermore, if unnecessarily employed, caching can eventually decrease performance rather than improve it.

\emph{\textbf{Document and report measurements.}} \adhoctuningperformanceevidence To compare and reproduce performance tests employed, it is important to document the setup used to perform the application analysis. It includes modules enabled, particular configurations and hardware settings.

\emph{\textbf{Consider using supporting libraries and frameworks.}} \externaldependecyevidence and \externalcomponentbenefitsevidence Supporting libraries and frameworks can raise the level of abstraction of cache implementation and provide useful features. In addition, they can scale up in a much easier and faster way than application-wide solutions.

\emph{\textbf{Tune default configurations.}} \externaldependecyevidence, \uncertaintyexternalcomponentevidence, \sizedefinitionevidence and \evictiondefinitionevidence Default configurations provided by external components serve as a start point. However, because they are generic and may not fit the application specificities, other configurations must be evaluated and possibly adopted.

\emph{\textbf{Use of transparent caching components.}} \externaldependecyevidence and \uncertaintyexternalcomponentevidence The use of transparent caching solutions to address bottlenecks outside the application boundaries such as databases, final HTML pages or fragments, and static assets can provide fast results. These solutions do not explore application specificities, but can still provide performance benefits for typical usage scenarios.

All the proposed guidelines are summarized in Table~\ref{tab:guidelines} with the respective name, associated evidence and identification.

\begin{table}
\centering
\renewcommand\tabcolsep{1mm}
\caption{Caching Guidelines Derived from the Study.}
\label{tab:guidelines}
\begin{tabular}{|p{3.5cm}|p{3.8cm}|p{.9cm}|}
\hline
\textbf{Guideline} & \textbf{Associated Evidence}& \textbf{ID}  \\ \hline
Evaluate different abstraction levels to cache & \thereisnorightplaceevidence and \cachinginboundariesevidence & DG-01 \\ \hline
Stack caching layers & \thereisnorightplaceevidence and \cachinglayerscriterionevidence & DG-02 \\ \hline
Separate dynamic from static data & \datachangeabilitycriterionvidence and \timesensitivecriterionevidence & DG-03 \\ \hline
Evaluate application boundaries & \cachinglayerscriterionevidence and \cachinginboundariesevidence & DG-04 \\ \hline
Specify selection criteria & \uncertaindesignevidence, \timesensitivecriterionevidence, \datachangeabilitycriterionvidence, \expensivedatacriterionevidence, \sizeofdatacriterionevidence and \sizeofcachecriterionevidence & DG-05 \\ \hline
Evaluate staleness and lifetime of cached data & \timesensitivecriterionevidence, \datachangeabilitycriterionvidence and \ensuringconsistencyevidence & DG-06 \\ \hline
Avoid caching per-user data & \userdatacriterionevidence and \sizedefinitionevidence& DG-07 \\ \hline
Avoid caching volatile data & \timesensitivecriterionevidence and \datachangeabilitycriterionvidence& DG-08 \\ \hline
Do not discard small improvements & \datachangeabilitycriterionvidence, \smallimprovementsevidence and \thereisnorightplaceevidence& DG-09 \\ \hline

Keep the cache API simple & \maintenanceproblemevidence, \codereuseevidence, \compleximplementationevidence and \externaldependecyevidence& IG-01 \\ \hline
Define naming conventions & \namingconventionevidence and \complexnamingevidence& IG-02 \\ \hline
Perform cache actions asynchronously & \backgroundprocessevidence& IG-03 \\ \hline
Do not use cache as data storage & \evictiondefinitionevidence& IG-04 \\ \hline

Perform measurements & \uncertaindesignevidence, \smallimprovementsevidence, \sizedefinitionevidence, \tuningperformanceevidence and \adhoctuningperformanceevidence& MG-01 \\ \hline
Document and report measurements & \adhoctuningperformanceevidence & MG-02 \\ \hline
Consider using supporting libraries and frameworks & \externaldependecyevidence and \externalcomponentbenefitsevidence & MG-03 \\ \hline
Tune default configurations & \externaldependecyevidence, \uncertaintyexternalcomponentevidence, \sizedefinitionevidence and \evictiondefinitionevidence & MG-04 \\ \hline
Use of transparent caching components & \externaldependecyevidence and \uncertaintyexternalcomponentevidence& MG-05 \\ \hline
\end{tabular}
\end{table}

\subsection{Patterns}

Based on our study, we derived caching patterns, which can be used by developers to help them design, implement and manage cache. They can be used in combination with our guidelines. Moreover, we identified the components these patterns must have, which comprise a template for a caching pattern catalog. These components are (i) a \emph{classification}, (ii) the pattern \emph{intent}, (iii) the \emph{problem} involved, (iv) the \emph{solution} proposed, and (v) an \emph{example}.

The proposed patterns are summarized in Table~\ref{tab:patterns} along with the possible combination with guidelines. We, in this paper, limit ourselves to present in detail one of our patterns, namely the \emph{Cacheability Pattern}, in Table~\ref{tab:cacheabilitypattern}, as an example. The complete description of the remaining patterns derived from our study is available elsewhere~\cite{CachingPatterns2016}.

\begin{table}
\centering
\renewcommand\tabcolsep{1mm}
\caption{Caching Pattern Classification.}
\label{tab:patterns}
\begin{tabular}{|p{1.75cm}|p{1.95cm}|p{2.9cm}|p{1.5cm}|}
\hline
\textbf{Pattern} & \textbf{Classification} & \textbf{Intent} & \textbf{Associated Guidelines}\\ \hline
Asynchronous Loading & Implementation & Design a mediator to asynchronously deal with caching. & IG-03, MG-03 \\ \hline

Cacheability & Design & Provide an intuitive process to decide whether to cache or not particular data. & DG-01, DG-03, DG-05, DG-07, DG-08, DG-09 \\ \hline

Data{\par}Expiration & Design and{\par}Maintenance & Given cacheable content, provide an intuitive process to choose a consistency management approach based on data specificities. & DG-06, DG-08, MG-03, MG-04 \\ \hline

Name Assignment & Implementation & Ensure a unique key and keep track of the content cached. & IG-02 \\ \hline
\end{tabular}
\end{table}

\begin{table*}
\centering
\caption{Cacheability Pattern.}
\label{tab:cacheabilitypattern}
\begin{tabular}{|p{16cm}|}
\hline
\textbf{Classification:} Design          \\ \hline

\textbf{Intent:} provide an intuitive process to decide whether to cache or not particular data.  \\ \hline

\textbf{Problem:} cache has limited size, so it is important to use the available space to cache data that maximizes the benefits provided to the application. Otherwise, it can end up reducing application performance instead of improving it, consuming more cache memory and at the same time suffering from cache misses, where the data is not getting served from cache but is fetched from the source.  \\ \hline

\textbf{Solution:} even though there are many criteria that contribute for identifying the level of data cacheability, there is a subset that would confirm this decision regardless of the values of the other criteria. Changeability is the first criterion that should be analyzed while selecting cacheable data, then usage frequency, shareability, retrieval complexity, and cache properties should be considered.

Figure 6 expresses a flowchart of the reasoning process to decide whether to cache data, based on the observation of data and cache properties. All criteria are tightly related to the application specificities and should be specified by the developer.

\begin{center}
\includegraphics[scale=0.35]{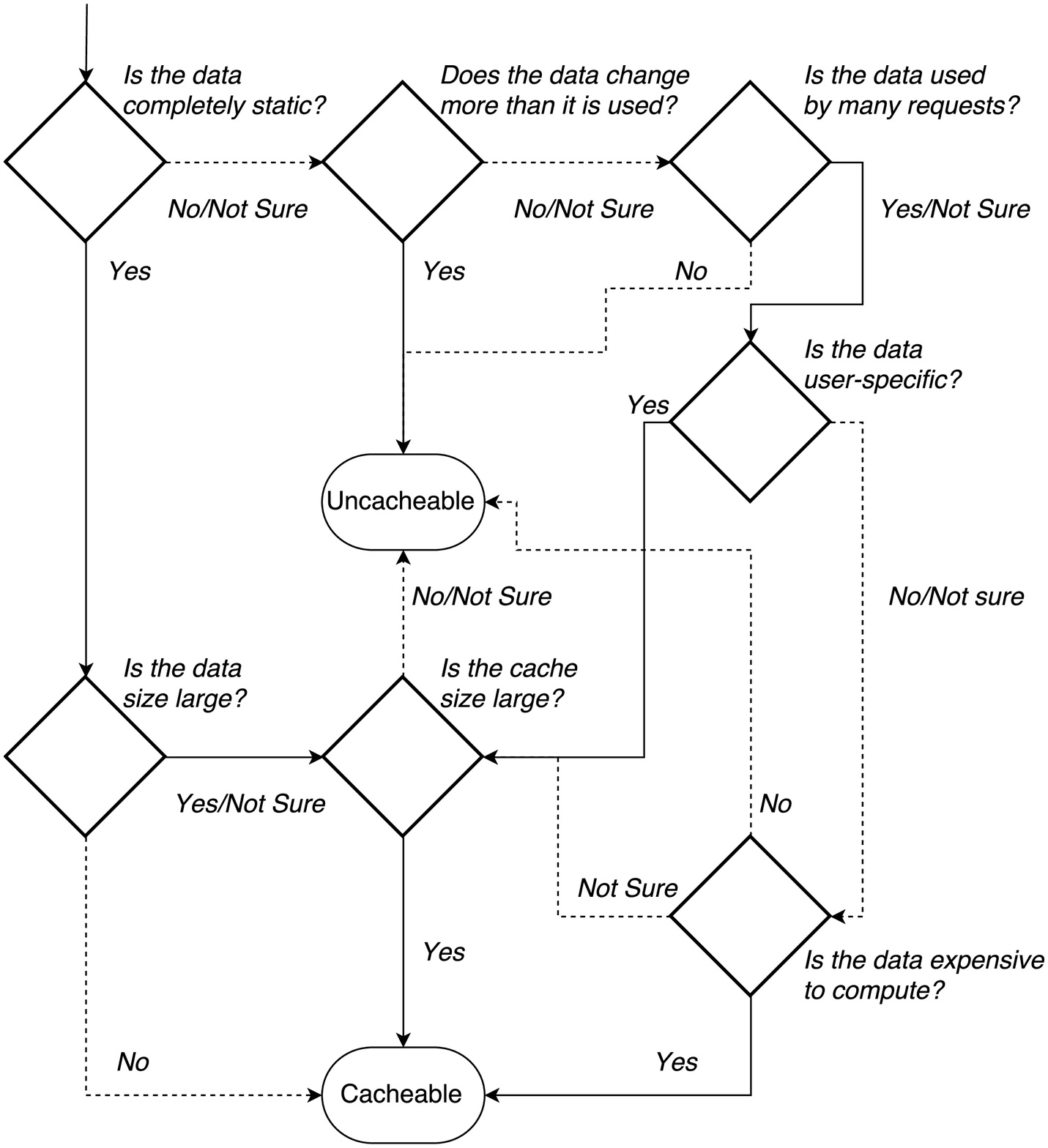}
\label{fig:cachebility}
\captionof{figure}{Cacheability flowchart: intuitive process to decide whether to cache or not particular data.}
\end{center}

\emph{Rules of thumb:}

(a) Despite being frequently used, user-specific data are not shareable and may not bring the benefit of caching, being usually avoided by developers. In this case, a specific session component is used to keep and retrieve user sessions.

(b) If the data changes frequently, it should not be immediately discarded from cache. An evaluation of the performance benefits of caching against the cost of building the cache should be done. Caching frequently changing data can provide benefits if slightly stale data is allowed.

(c) Expensive spots (when much processing is required to retrieve or create data) are bottlenecks that directly affect application performance and should be cached, even though it can increase complexity and responsibilities to deal with. Methods with high latency or that consists of a large call stack are some examples of this situation and opportunities for caching.

In addition, we list content properties that should be avoided, which do not convey the influence factors in a good way and lead to problems such as cache trashing.

(a) User-specific data. Avoid caching content that varies depending on the particularities of the request, unless weak consistency is acceptable. Otherwise, the cache can end up being fulfilled with small and less beneficial objects. As result, the caching component achieves its maximum capacity earlier and is flushed or replaced many times in a brief period, which is cache thrashing.

(b) Highly time-sensitive data. Content that changes more than is used should not be cached given that it will not take advantage from caching. The cost of implementing and designing an efficient consistency policy may not be compensate.

(c) Large-sized objects. Unless the size of the cache is large enough, do not cache large objects, it will probably result in a cache trashing problem, where the caching component is flushed or replaced many times in a short period. \\ \hline

\textbf{Example:} we list some typical scenarios where data should be cached and also give explanations based on the criteria presented.

(a) Headlines. In most cases, headlines are shared by multiple users and updated infrequently.

(b) Dashboards. Usually, much data need to be gathered across several application modules and manipulated to build a summarized information about the application.

(c) Catalogs. Catalogs need to be updated at specific intervals, are shared across the application, and manipulated before sending the content to the client.

(d) Metadata/configuration. Settings that do not frequently change, such as country/state lists, external resource addresses, logic/branching settings and tax definitions.

(e) Historical datasets for reports. Costly to retrieve or create and does not need to change frequently. \\ \hline

\end{tabular}
\end{table*}

\section{Threats to validity} \label{sec:threats}

We now analyze the possible threats to the validity of this study, and how we mitigated them. Researcher bias is a typical threat to qualitative research because the results are subject to the researcher interpretation of the data. In our study, prior knowledge might have influenced results. In order to avoid this, we followed a systematic analysis of our data, and conclusions are all founded on these data. Moreover, cross-checks were performed using our different sources of evidence.

Another source of bias in this study is the process adopted to select applications and the number of applications analyzed. We mitigated this problem by specifying criteria (described in Section~\ref{sec:targetsystems}) that allowed us to select representative systems with different characteristics to be part of this study. Moreover, we observed that our codes and patterns emerged after analyzing part of our systems and, in the remaining ones, we identified only recurrences of codes and patterns. This is an evidence that the investigation of a larger number of systems would not bring more information. Furthermore, given that the goal of qualitative research is to understand a particular event, rather than to provide a general description of a large sample of a population, selecting few representative applications is sufficient for our study.

The existence of such plateau reached after the analysis of the fourth application also gives evidence that the results may be widely applied to other web applications. However, we acknowledge that the selected applications may have influenced the results of this study as well as the order in which they were analyzed, thus being an external threat to the validity of this study. We mitigated this threat by selecting a sample of applications with different domains, programming languages, technologies, sizes and business models, and the use of a random order to analyze them.

Despite caching is associated with performance concerns and requirements, the focus of this study is on software engineering concerns of application-level caching, approaching design, implementation, and maintenance from a developer point of view. It is important to highlight that our goal is to capture the state-of-practice of application-level caching, by characterizing and deriving caching patterns. Consequently, we assume that the investigated applications have adequate caching implementations given that caching was introduced in previous releases and was already debugged and improved by developers, i.e.\ we assume that the design and implementation are all efficient; otherwise, there would be open issues in issue trackers to be resolved by developers. Therefore, another threat in our research is related to the possibility that we observed inadequately designed and implemented caching. Furthermore, the quality or performance improvement of design and implementation of application-level caching could be better explored in future work.

Finally, most of the web applications satisfy initial performance and scalability requirements using traditional forms of caching. The need for application-level caching usually appears after the application is released, with an increasing user demand. Due to this, all applications selected in our study did not adopt application-level caching since their conception---no application implemented with application-level caching in its first release was found. Consequently, it is not possible to assess if the overhead of adding cache to an application in later stages is higher than doing so during its initial development. For all analyzed systems, performed complexity or usage analysis led to requests for improvements, as systems became larger. Consequently, developers had to refactor the application to insert caching logic. This is thus another external threat to validity, given that our results cannot be generalized to systems that use application-level caching since they are conceived.

\section{Discussion and Conclusion}
\label{sec:discussion-conclusion}

Application-level caching has been increasingly used in the development of web applications, in order to improve their response time given that they are becoming more complex and dealing with larger amounts of data over time. Caching has been used in different locations, such as proxy servers, often as seamless components. Application-level caching allows caching additional content taking into account application specificities not captured by off-the-shelf components. However, there is limited guidance to design, implement and manage application-level caching, which is often implemented in an \emph{ad-hoc} way. In this paper, we presented a qualitative study performed to understand how developers approach application-level caching. The study consisted of the selection ten web applications and investigation of caching-related aspects, namely design, implementation and maintenance practices.

We observed a higher number of categories, which are classes of observations made based on the analysis of these applications, associated with caching design and implementation than those associated with maintenance. Furthermore, the number of occurrences of each category, design and implementation categories also have higher numbers. This phenomenon was expected since the most representative portion of our qualitative data consists of source code and issues. Moreover, simple maintenance tasks and configurations are already executed and provided by external components, being commonly adopted. However, the number of occurrences do not reflect the importance of a category.

The high number of occurrences related to \emph{ensuring consistency} refers to the expiration process of cached content, which requires extra reasoning from developers, because they should track which changes cause data content to become outdated, and be aware for how long the cache can provide stale data, in case the data source has been updated. In fact, consistency approaches have been widely investigated~\cite{Ports2010,Gupta2011}, and the typical way of dealing with it is to analyze data dependency, from which conditions and constraints for consistency can be derived.

Many emerged categories, such as \emph{Bugs and Issues}, \emph{Technical Debt Concerns} and \emph{Complex Design and Implementation}, indicate the additional burden placed on developers while designing and implementing application-level caching, which can lead to a significant amount of time and effort added to software projects. Therefore, ideally, opportunities for caching should be identified as early as possible during the application design, avoiding the problems of employing cache only in late stages of the development cycle or even after it is in production, as an emergency solution. Furthermore, we observed that faster results can be achieved when a caching component is not implemented from scratch, but when libraries and frameworks are adopted to support cache implementation.

Although a well designed and implemented cache can achieve desired non-functional requirements, its performance may decay over time due to changes in the application domain, workload characteristics and access patterns. Therefore, achieving acceptable application performance requires constantly tuning cache decisions, which implies extra time dedicated to maintenance~\cite{Radhakrishnan2004}. 

From all analyzed applications, we observed that none of them use a proactive approach to cache content. Content is always cached after it requested (i.e.\ reactive approach) and, as a consequence, the first request always results in a cache miss. Due to this, prefetching techniques can be used in order to populate the cache and prevent misses by predicting and caching data that will potentially be requested in the future. It can be based on heuristics, usage observations or even with the use of complex prediction algorithms~\cite{Podlipnig2003,Domenech2006,Ali2011,Pal2014a}. However, the design and implementation of a reactive cache component already requires significant effort and reasoning to be properly done, and a proactive approach increases the complexity of the caching solution even more.

Based on the results of our study, guidelines and patterns were derived to provide guidance to developers to deal with application-level caching. The former are high-level directions to develop caching solutions, and also serve as a checklist of points that must be analyzed while designing and implementing a caching component. The latter capture the reasoning used by developers to build a caching component and integrating it with a web application, providing a systematic way to develop a caching solution. Regardless of the use of our guidelines or patterns (or any other reusable component), it is important to determine exactly whether a particular caching approach is performing adequately. Therefore, it is essential to perform an application profiling, by measuring performance both with and without caching.

Our guidelines and patterns can be used as foundation to design, implement and manage a caching component for a particular application as well as to develop an application-independent caching component or framework. Future work involves the use of the guidance derived from our qualitative research to offer solutions to raise the abstraction level of application-level caching and, by distinguishing domain-neutral from domain-specific caching aspects, automate the reasoning captured in our patterns, providing a better overall experience with caching for developers.

 \ifCLASSOPTIONcompsoc
   \section*{Acknowledgments}
 \else
   \section*{Acknowledgment}
 \fi
Jhonny Mertz would like to thank CNPq grant 131550/2015-2. Ingrid Nunes thanks for research grants CNPq ref. 303232/2015-3, CAPES ref. 7619-15-4, and Alexander von Humboldt, ref. BRA 1184533 HFSTCAPES-P. We would also like to thank the editor of this journal and anonymous reviewers who have provided helpful comments and suggestions on the refinement of the paper.

%
%






\newpage

\begin{IEEEbiography}[{\includegraphics[width=1in,height=1.25in,clip,keepaspectratio]{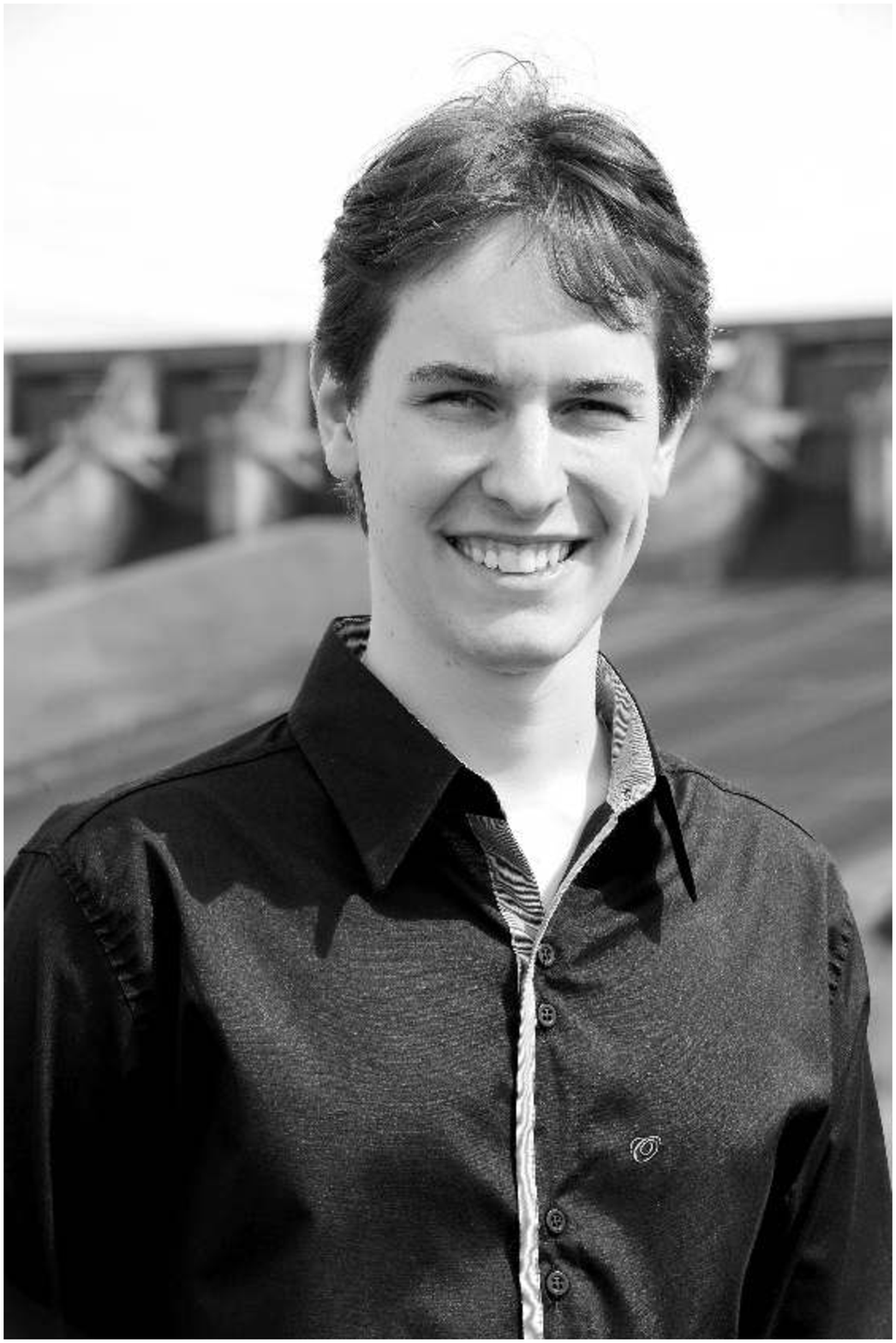}}]{Jhonny Mertz}
is an MSc student in computer science at the Federal University of Rio Grande do Sul (UFRGS), in Brazil. He obtained his BSc from the Universidade Estadual do Oeste do Paran\'{a} (UNIOESTE), Brazil, in 2012. His research interests include software maintenance and evolution, application-level caching and autonomic computing. See http://www.inf.ufrgs.br/$\sim$jmamertz for more information.
\end{IEEEbiography}

\begin{IEEEbiography}[{\includegraphics[width=1in,height=1.25in,clip,keepaspectratio]{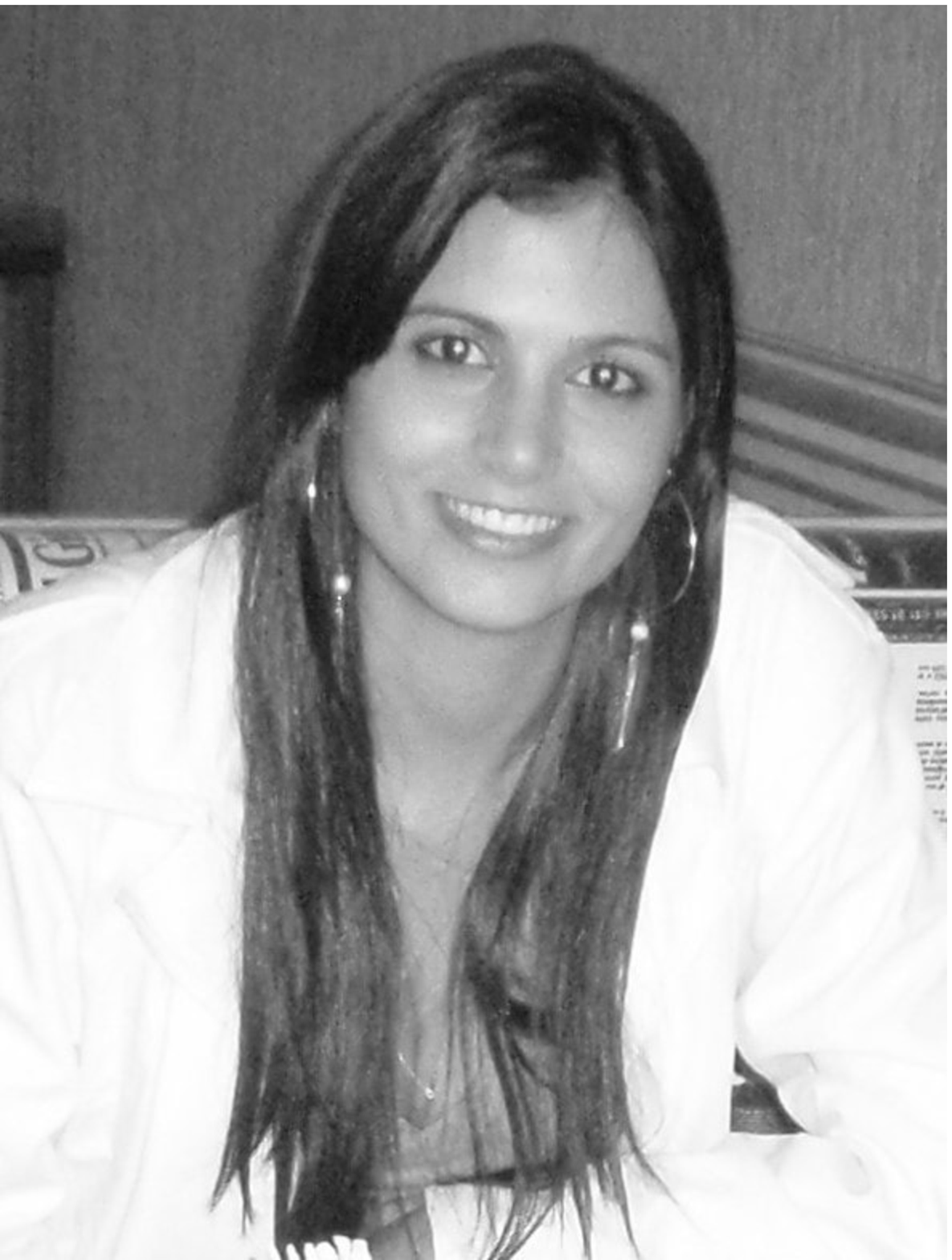}}]{Ingrid Nunes}
is an Associate Professor at the Institute of Informatics, Universidade Federal do Rio Grande do Sul (UFRGS), Brazil, currently in a sabbatical year at TU Dortmund in Germany.  She obtained her PhD in Informatics at the Pontifical Catholic University of Rio de Janeiro (PUC-Rio), Brazil. Her PhD was in cooperation with King's College London (UK) and University of Waterloo (Canada). She is the head of the Prosoft research group, and her main research areas are software maintenance and evolution and agent-oriented software engineering.
\end{IEEEbiography}

\vfill







\end{document}